\documentclass[longauth]{aaEC}

\usepackage{graphicx}
\usepackage{natbib}
\usepackage{scalerel}

\usepackage[table]{xcolor}

\usepackage{txfonts}
\usepackage[pdfencoding=auto,psdextra]{hyperref}
\hypersetup{
    colorlinks=true,
    linkcolor=blue,
    filecolor=magenta,  urlcolor=blue,
    citecolor=blue
}
\makeatletter
\renewcommand*\aa@pageof{, page \thepage{} of \pageref*{LastPage}}
\makeatother

\newcommand{\gsim}{\raisebox{-0.13cm}{~\shortstack{$>$ \\[-0.07cm] $\sim$}}~}

\usepackage[utf8]{inputenc}

\usepackage[switch, modulo]{lineno}

\usepackage{euclid}

\begin{document}
%
% Put the title of your paper here:
%
\title{Euclid Quick Data Release (Q1)}
\subtitle{Identification of massive galaxy candidates at the end of the Epoch of Reionisation}    

%% please do not edit the author list -- contact ECEB Bureau for changes
% \newcommand{\orcid}[1]{} %% define as link to https://orcid.org/#1 if needed
% \author{\normalsize F.~Author$^{1}$\thanks{Corresponding author, \email{author@institute.mail}}, S.~Writer$^{2,3}$, A.~Third$^{3}$, F.~Plotter$^{4}$}

% \institute{$^{1}$ Dark Energy Institute, Metropolis, 80th Backyard
%   St., Freedonia\\
% $^{2}$ Center for Large-Scale Structure, University of Somewhere,
% Trembley 48, GoK 23456, Backland\\
% $^{3}$ Institute for Fundamental Research, Rainbow University, 812
% Clear Sky Road, 83456 Rainbow, Wonderland  \\
% $^{4}$ Institute for Astrophysics and Cosmology, UC Gilroy, Major
% Garlic St., Gilroy, CA 93712, USA\\}
%%%% please do not edit the author list -- contact ECEB Bureau for changes
%%%% Version Monday 27th of October 2025 05:49:22 AM UT		

%%%% Version Wednesday 12th of November 2025 09:17:25 AM UT												
%%%% Please do not edit the author list -- contact ECEB Bureau for changes
\newcommand{\orcid}[1]{} %% if already defined in aa.cls: comment, or use renewcommand			   
\author{Euclid Collaboration: R.~Navarro-Carrera\orcid{0000-0001-6066-4624}\thanks{\email{navarro@astro.rug.nl}}\inst{\ref{aff1}}
\and K.~I.~Caputi\orcid{0000-0001-8183-1460}\inst{\ref{aff1},\ref{aff2}}
\and C.~J.~R.~McPartland\orcid{0000-0003-0639-025X}\inst{\ref{aff3},\ref{aff4}}
\and J.~R.~Weaver\orcid{0000-0003-1614-196X}\inst{\ref{aff5}}
\and D.~B.~Sanders\orcid{0000-0002-1233-9998}\inst{\ref{aff6}}
\and G.~Desprez\orcid{0000-0001-8325-1742}\inst{\ref{aff1}}
\and A.~A.~Tumborang\orcid{0009-0008-3005-0435}\inst{\ref{aff1}}
\and A.~Biviano\orcid{0000-0002-0857-0732}\inst{\ref{aff7},\ref{aff8}}
\and C.~J.~Conselice\orcid{0000-0003-1949-7638}\inst{\ref{aff9}}
\and Y.~Fu\orcid{0000-0002-0759-0504}\inst{\ref{aff10},\ref{aff1}}
\and G.~Girardi\orcid{0009-0005-6156-4066}\inst{\ref{aff11},\ref{aff12}}
\and V.~Le~Brun\orcid{0000-0002-5027-1939}\inst{\ref{aff13}}
\and C.~C.~Lovell\orcid{0000-0001-7964-5933}\inst{\ref{aff14},\ref{aff15}}
\and G.~Rodighiero\orcid{0000-0002-9415-2296}\inst{\ref{aff11},\ref{aff12}}
\and J.~Schaye\orcid{0000-0002-0668-5560}\inst{\ref{aff10}}
\and R.~G.~Varadaraj\orcid{0009-0006-9953-6471}\inst{\ref{aff16}}
\and S.~M.~Wilkins\orcid{0000-0003-3903-6935}\inst{\ref{aff17}}
\and G.~Zamorani\orcid{0000-0002-2318-301X}\inst{\ref{aff18}}
\and K.~Jahnke\orcid{0000-0003-3804-2137}\inst{\ref{aff19}}
\and D.~Scott\orcid{0000-0002-6878-9840}\inst{\ref{aff20}}
\and M.~Siudek\orcid{0000-0002-2949-2155}\inst{\ref{aff21},\ref{aff22}}
\and F.~Shankar\orcid{0000-0001-8973-5051}\inst{\ref{aff23}}
\and J.~G.~Sorce\orcid{0000-0002-2307-2432}\inst{\ref{aff24},\ref{aff25}}
\and F.~Tarsitano\orcid{0000-0002-5919-0238}\inst{\ref{aff26},\ref{aff27}}
\and A.~Amara\inst{\ref{aff28}}
\and S.~Andreon\orcid{0000-0002-2041-8784}\inst{\ref{aff29}}
\and N.~Auricchio\orcid{0000-0003-4444-8651}\inst{\ref{aff18}}
\and C.~Baccigalupi\orcid{0000-0002-8211-1630}\inst{\ref{aff8},\ref{aff7},\ref{aff30},\ref{aff31}}
\and M.~Baldi\orcid{0000-0003-4145-1943}\inst{\ref{aff32},\ref{aff18},\ref{aff33}}
\and A.~Balestra\orcid{0000-0002-6967-261X}\inst{\ref{aff12}}
\and S.~Bardelli\orcid{0000-0002-8900-0298}\inst{\ref{aff18}}
\and P.~Battaglia\orcid{0000-0002-7337-5909}\inst{\ref{aff18}}
\and E.~Branchini\orcid{0000-0002-0808-6908}\inst{\ref{aff34},\ref{aff35},\ref{aff29}}
\and M.~Brescia\orcid{0000-0001-9506-5680}\inst{\ref{aff36},\ref{aff37}}
\and J.~Brinchmann\orcid{0000-0003-4359-8797}\inst{\ref{aff38},\ref{aff39},\ref{aff40}}
\and G.~Ca\~nas-Herrera\orcid{0000-0003-2796-2149}\inst{\ref{aff41},\ref{aff10}}
\and V.~Capobianco\orcid{0000-0002-3309-7692}\inst{\ref{aff42}}
\and C.~Carbone\orcid{0000-0003-0125-3563}\inst{\ref{aff43}}
\and J.~Carretero\orcid{0000-0002-3130-0204}\inst{\ref{aff44},\ref{aff45}}
\and M.~Castellano\orcid{0000-0001-9875-8263}\inst{\ref{aff46}}
\and G.~Castignani\orcid{0000-0001-6831-0687}\inst{\ref{aff18}}
\and S.~Cavuoti\orcid{0000-0002-3787-4196}\inst{\ref{aff37},\ref{aff47}}
\and K.~C.~Chambers\orcid{0000-0001-6965-7789}\inst{\ref{aff6}}
\and A.~Cimatti\inst{\ref{aff48}}
\and C.~Colodro-Conde\inst{\ref{aff49}}
\and G.~Congedo\orcid{0000-0003-2508-0046}\inst{\ref{aff50}}
\and L.~Conversi\orcid{0000-0002-6710-8476}\inst{\ref{aff51},\ref{aff52}}
\and Y.~Copin\orcid{0000-0002-5317-7518}\inst{\ref{aff53}}
\and F.~Courbin\orcid{0000-0003-0758-6510}\inst{\ref{aff54},\ref{aff55},\ref{aff56}}
\and H.~M.~Courtois\orcid{0000-0003-0509-1776}\inst{\ref{aff57}}
\and M.~Cropper\orcid{0000-0003-4571-9468}\inst{\ref{aff58}}
\and A.~Da~Silva\orcid{0000-0002-6385-1609}\inst{\ref{aff59},\ref{aff60}}
\and H.~Degaudenzi\orcid{0000-0002-5887-6799}\inst{\ref{aff27}}
\and G.~De~Lucia\orcid{0000-0002-6220-9104}\inst{\ref{aff7}}
\and A.~M.~Di~Giorgio\orcid{0000-0002-4767-2360}\inst{\ref{aff61}}
\and H.~Dole\orcid{0000-0002-9767-3839}\inst{\ref{aff25}}
\and F.~Dubath\orcid{0000-0002-6533-2810}\inst{\ref{aff27}}
\and C.~A.~J.~Duncan\orcid{0009-0003-3573-0791}\inst{\ref{aff50}}
\and X.~Dupac\inst{\ref{aff52}}
\and S.~Dusini\orcid{0000-0002-1128-0664}\inst{\ref{aff62}}
\and S.~Escoffier\orcid{0000-0002-2847-7498}\inst{\ref{aff63}}
\and M.~Farina\orcid{0000-0002-3089-7846}\inst{\ref{aff61}}
\and R.~Farinelli\inst{\ref{aff18}}
\and F.~Faustini\orcid{0000-0001-6274-5145}\inst{\ref{aff46},\ref{aff64}}
\and S.~Ferriol\inst{\ref{aff53}}
\and F.~Finelli\orcid{0000-0002-6694-3269}\inst{\ref{aff18},\ref{aff65}}
\and M.~Frailis\orcid{0000-0002-7400-2135}\inst{\ref{aff7}}
\and E.~Franceschi\orcid{0000-0002-0585-6591}\inst{\ref{aff18}}
\and M.~Fumana\orcid{0000-0001-6787-5950}\inst{\ref{aff43}}
\and S.~Galeotta\orcid{0000-0002-3748-5115}\inst{\ref{aff7}}
\and K.~George\orcid{0000-0002-1734-8455}\inst{\ref{aff66}}
\and B.~Gillis\orcid{0000-0002-4478-1270}\inst{\ref{aff50}}
\and C.~Giocoli\orcid{0000-0002-9590-7961}\inst{\ref{aff18},\ref{aff33}}
\and J.~Gracia-Carpio\inst{\ref{aff67}}
\and A.~Grazian\orcid{0000-0002-5688-0663}\inst{\ref{aff12}}
\and F.~Grupp\inst{\ref{aff67},\ref{aff68}}
\and S.~V.~H.~Haugan\orcid{0000-0001-9648-7260}\inst{\ref{aff69}}
\and H.~Hoekstra\orcid{0000-0002-0641-3231}\inst{\ref{aff10}}
\and W.~Holmes\inst{\ref{aff70}}
\and I.~M.~Hook\orcid{0000-0002-2960-978X}\inst{\ref{aff71}}
\and F.~Hormuth\inst{\ref{aff72}}
\and A.~Hornstrup\orcid{0000-0002-3363-0936}\inst{\ref{aff73},\ref{aff3}}
\and M.~Jhabvala\inst{\ref{aff74}}
\and B.~Joachimi\orcid{0000-0001-7494-1303}\inst{\ref{aff75}}
\and E.~Keih\"anen\orcid{0000-0003-1804-7715}\inst{\ref{aff76}}
\and S.~Kermiche\orcid{0000-0002-0302-5735}\inst{\ref{aff63}}
\and A.~Kiessling\orcid{0000-0002-2590-1273}\inst{\ref{aff70}}
\and B.~Kubik\orcid{0009-0006-5823-4880}\inst{\ref{aff53}}
\and M.~K\"ummel\orcid{0000-0003-2791-2117}\inst{\ref{aff68}}
\and M.~Kunz\orcid{0000-0002-3052-7394}\inst{\ref{aff77}}
\and H.~Kurki-Suonio\orcid{0000-0002-4618-3063}\inst{\ref{aff78},\ref{aff79}}
\and A.~M.~C.~Le~Brun\orcid{0000-0002-0936-4594}\inst{\ref{aff80}}
\and S.~Ligori\orcid{0000-0003-4172-4606}\inst{\ref{aff42}}
\and P.~B.~Lilje\orcid{0000-0003-4324-7794}\inst{\ref{aff69}}
\and V.~Lindholm\orcid{0000-0003-2317-5471}\inst{\ref{aff78},\ref{aff79}}
\and I.~Lloro\orcid{0000-0001-5966-1434}\inst{\ref{aff81}}
\and G.~Mainetti\orcid{0000-0003-2384-2377}\inst{\ref{aff82}}
\and D.~Maino\inst{\ref{aff83},\ref{aff43},\ref{aff84}}
\and E.~Maiorano\orcid{0000-0003-2593-4355}\inst{\ref{aff18}}
\and O.~Mansutti\orcid{0000-0001-5758-4658}\inst{\ref{aff7}}
\and O.~Marggraf\orcid{0000-0001-7242-3852}\inst{\ref{aff85}}
\and M.~Martinelli\orcid{0000-0002-6943-7732}\inst{\ref{aff46},\ref{aff86}}
\and N.~Martinet\orcid{0000-0003-2786-7790}\inst{\ref{aff13}}
\and F.~Marulli\orcid{0000-0002-8850-0303}\inst{\ref{aff87},\ref{aff18},\ref{aff33}}
\and R.~J.~Massey\orcid{0000-0002-6085-3780}\inst{\ref{aff88}}
\and E.~Medinaceli\orcid{0000-0002-4040-7783}\inst{\ref{aff18}}
\and S.~Mei\orcid{0000-0002-2849-559X}\inst{\ref{aff89},\ref{aff90}}
\and M.~Melchior\inst{\ref{aff91}}
\and Y.~Mellier\inst{\ref{aff92},\ref{aff93}}
\and M.~Meneghetti\orcid{0000-0003-1225-7084}\inst{\ref{aff18},\ref{aff33}}
\and E.~Merlin\orcid{0000-0001-6870-8900}\inst{\ref{aff46}}
\and G.~Meylan\inst{\ref{aff94}}
\and A.~Mora\orcid{0000-0002-1922-8529}\inst{\ref{aff95}}
\and M.~Moresco\orcid{0000-0002-7616-7136}\inst{\ref{aff87},\ref{aff18}}
\and L.~Moscardini\orcid{0000-0002-3473-6716}\inst{\ref{aff87},\ref{aff18},\ref{aff33}}
\and R.~Nakajima\orcid{0009-0009-1213-7040}\inst{\ref{aff85}}
\and C.~Neissner\orcid{0000-0001-8524-4968}\inst{\ref{aff96},\ref{aff45}}
\and R.~C.~Nichol\orcid{0000-0003-0939-6518}\inst{\ref{aff28}}
\and S.-M.~Niemi\orcid{0009-0005-0247-0086}\inst{\ref{aff41}}
\and C.~Padilla\orcid{0000-0001-7951-0166}\inst{\ref{aff96}}
\and S.~Paltani\orcid{0000-0002-8108-9179}\inst{\ref{aff27}}
\and F.~Pasian\orcid{0000-0002-4869-3227}\inst{\ref{aff7}}
\and K.~Pedersen\inst{\ref{aff97}}
\and W.~J.~Percival\orcid{0000-0002-0644-5727}\inst{\ref{aff98},\ref{aff99},\ref{aff100}}
\and V.~Pettorino\orcid{0000-0002-4203-9320}\inst{\ref{aff41}}
\and S.~Pires\orcid{0000-0002-0249-2104}\inst{\ref{aff101}}
\and G.~Polenta\orcid{0000-0003-4067-9196}\inst{\ref{aff64}}
\and M.~Poncet\inst{\ref{aff102}}
\and L.~A.~Popa\inst{\ref{aff103}}
\and L.~Pozzetti\orcid{0000-0001-7085-0412}\inst{\ref{aff18}}
\and A.~Renzi\orcid{0000-0001-9856-1970}\inst{\ref{aff11},\ref{aff62}}
\and J.~Rhodes\orcid{0000-0002-4485-8549}\inst{\ref{aff70}}
\and G.~Riccio\inst{\ref{aff37}}
\and E.~Romelli\orcid{0000-0003-3069-9222}\inst{\ref{aff7}}
\and M.~Roncarelli\orcid{0000-0001-9587-7822}\inst{\ref{aff18}}
\and R.~Saglia\orcid{0000-0003-0378-7032}\inst{\ref{aff68},\ref{aff67}}
\and Z.~Sakr\orcid{0000-0002-4823-3757}\inst{\ref{aff104},\ref{aff105},\ref{aff106}}
\and D.~Sapone\orcid{0000-0001-7089-4503}\inst{\ref{aff107}}
\and M.~Schirmer\orcid{0000-0003-2568-9994}\inst{\ref{aff19}}
\and P.~Schneider\orcid{0000-0001-8561-2679}\inst{\ref{aff85}}
\and T.~Schrabback\orcid{0000-0002-6987-7834}\inst{\ref{aff108}}
\and A.~Secroun\orcid{0000-0003-0505-3710}\inst{\ref{aff63}}
\and G.~Seidel\orcid{0000-0003-2907-353X}\inst{\ref{aff19}}
\and S.~Serrano\orcid{0000-0002-0211-2861}\inst{\ref{aff109},\ref{aff110},\ref{aff22}}
\and P.~Simon\inst{\ref{aff85}}
\and C.~Sirignano\orcid{0000-0002-0995-7146}\inst{\ref{aff11},\ref{aff62}}
\and G.~Sirri\orcid{0000-0003-2626-2853}\inst{\ref{aff33}}
\and L.~Stanco\orcid{0000-0002-9706-5104}\inst{\ref{aff62}}
\and J.~Steinwagner\orcid{0000-0001-7443-1047}\inst{\ref{aff67}}
\and P.~Tallada-Cresp\'{i}\orcid{0000-0002-1336-8328}\inst{\ref{aff44},\ref{aff45}}
\and A.~N.~Taylor\inst{\ref{aff50}}
\and H.~I.~Teplitz\orcid{0000-0002-7064-5424}\inst{\ref{aff111}}
\and I.~Tereno\orcid{0000-0002-4537-6218}\inst{\ref{aff59},\ref{aff112}}
\and N.~Tessore\orcid{0000-0002-9696-7931}\inst{\ref{aff75},\ref{aff58}}
\and S.~Toft\orcid{0000-0003-3631-7176}\inst{\ref{aff2},\ref{aff4}}
\and R.~Toledo-Moreo\orcid{0000-0002-2997-4859}\inst{\ref{aff113}}
\and F.~Torradeflot\orcid{0000-0003-1160-1517}\inst{\ref{aff45},\ref{aff44}}
\and I.~Tutusaus\orcid{0000-0002-3199-0399}\inst{\ref{aff22},\ref{aff109},\ref{aff105}}
\and L.~Valenziano\orcid{0000-0002-1170-0104}\inst{\ref{aff18},\ref{aff65}}
\and J.~Valiviita\orcid{0000-0001-6225-3693}\inst{\ref{aff78},\ref{aff79}}
\and T.~Vassallo\orcid{0000-0001-6512-6358}\inst{\ref{aff7}}
\and A.~Veropalumbo\orcid{0000-0003-2387-1194}\inst{\ref{aff29},\ref{aff35},\ref{aff34}}
\and Y.~Wang\orcid{0000-0002-4749-2984}\inst{\ref{aff111}}
\and J.~Weller\orcid{0000-0002-8282-2010}\inst{\ref{aff68},\ref{aff67}}
\and E.~Zucca\orcid{0000-0002-5845-8132}\inst{\ref{aff18}}
\and M.~Ballardini\orcid{0000-0003-4481-3559}\inst{\ref{aff114},\ref{aff115},\ref{aff18}}
\and M.~Bolzonella\orcid{0000-0003-3278-4607}\inst{\ref{aff18}}
\and E.~Bozzo\orcid{0000-0002-8201-1525}\inst{\ref{aff27}}
\and C.~Burigana\orcid{0000-0002-3005-5796}\inst{\ref{aff116},\ref{aff65}}
\and R.~Cabanac\orcid{0000-0001-6679-2600}\inst{\ref{aff105}}
\and M.~Calabrese\orcid{0000-0002-2637-2422}\inst{\ref{aff117},\ref{aff43}}
\and A.~Cappi\inst{\ref{aff118},\ref{aff18}}
\and J.~A.~Escartin~Vigo\inst{\ref{aff67}}
\and L.~Gabarra\orcid{0000-0002-8486-8856}\inst{\ref{aff16}}
\and W.~G.~Hartley\inst{\ref{aff27}}
\and M.~Huertas-Company\orcid{0000-0002-1416-8483}\inst{\ref{aff49},\ref{aff21},\ref{aff119},\ref{aff120}}
\and R.~Maoli\orcid{0000-0002-6065-3025}\inst{\ref{aff121},\ref{aff46}}
\and J.~Mart\'{i}n-Fleitas\orcid{0000-0002-8594-569X}\inst{\ref{aff122}}
\and S.~Matthew\orcid{0000-0001-8448-1697}\inst{\ref{aff50}}
\and M.~Maturi\orcid{0000-0002-3517-2422}\inst{\ref{aff104},\ref{aff123}}
\and N.~Mauri\orcid{0000-0001-8196-1548}\inst{\ref{aff48},\ref{aff33}}
\and R.~B.~Metcalf\orcid{0000-0003-3167-2574}\inst{\ref{aff87},\ref{aff18}}
\and A.~Pezzotta\orcid{0000-0003-0726-2268}\inst{\ref{aff29}}
\and M.~P\"ontinen\orcid{0000-0001-5442-2530}\inst{\ref{aff78}}
\and C.~Porciani\orcid{0000-0002-7797-2508}\inst{\ref{aff85}}
\and I.~Risso\orcid{0000-0003-2525-7761}\inst{\ref{aff29},\ref{aff35}}
\and V.~Scottez\orcid{0009-0008-3864-940X}\inst{\ref{aff92},\ref{aff124}}
\and M.~Sereno\orcid{0000-0003-0302-0325}\inst{\ref{aff18},\ref{aff33}}
\and M.~Tenti\orcid{0000-0002-4254-5901}\inst{\ref{aff33}}
\and M.~Viel\orcid{0000-0002-2642-5707}\inst{\ref{aff8},\ref{aff7},\ref{aff31},\ref{aff30},\ref{aff125}}
\and M.~Wiesmann\orcid{0009-0000-8199-5860}\inst{\ref{aff69}}
\and Y.~Akrami\orcid{0000-0002-2407-7956}\inst{\ref{aff126},\ref{aff127}}
\and I.~T.~Andika\orcid{0000-0001-6102-9526}\inst{\ref{aff128},\ref{aff129}}
\and S.~Anselmi\orcid{0000-0002-3579-9583}\inst{\ref{aff62},\ref{aff11},\ref{aff130}}
\and M.~Archidiacono\orcid{0000-0003-4952-9012}\inst{\ref{aff83},\ref{aff84}}
\and F.~Atrio-Barandela\orcid{0000-0002-2130-2513}\inst{\ref{aff131}}
\and D.~Bertacca\orcid{0000-0002-2490-7139}\inst{\ref{aff11},\ref{aff12},\ref{aff62}}
\and M.~Bethermin\orcid{0000-0002-3915-2015}\inst{\ref{aff132}}
\and L.~Bisigello\orcid{0000-0003-0492-4924}\inst{\ref{aff12}}
\and A.~Blanchard\orcid{0000-0001-8555-9003}\inst{\ref{aff105}}
\and L.~Blot\orcid{0000-0002-9622-7167}\inst{\ref{aff133},\ref{aff80}}
\and M.~Bonici\orcid{0000-0002-8430-126X}\inst{\ref{aff98},\ref{aff43}}
\and S.~Borgani\orcid{0000-0001-6151-6439}\inst{\ref{aff134},\ref{aff8},\ref{aff7},\ref{aff30},\ref{aff125}}
\and M.~L.~Brown\orcid{0000-0002-0370-8077}\inst{\ref{aff9}}
\and S.~Bruton\orcid{0000-0002-6503-5218}\inst{\ref{aff135}}
\and A.~Calabro\orcid{0000-0003-2536-1614}\inst{\ref{aff46}}
\and B.~Camacho~Quevedo\orcid{0000-0002-8789-4232}\inst{\ref{aff8},\ref{aff31},\ref{aff7}}
\and F.~Caro\inst{\ref{aff46}}
\and C.~S.~Carvalho\inst{\ref{aff112}}
\and T.~Castro\orcid{0000-0002-6292-3228}\inst{\ref{aff7},\ref{aff30},\ref{aff8},\ref{aff125}}
\and F.~Cogato\orcid{0000-0003-4632-6113}\inst{\ref{aff87},\ref{aff18}}
\and S.~Conseil\orcid{0000-0002-3657-4191}\inst{\ref{aff53}}
\and T.~Contini\orcid{0000-0003-0275-938X}\inst{\ref{aff105}}
\and A.~R.~Cooray\orcid{0000-0002-3892-0190}\inst{\ref{aff136}}
\and O.~Cucciati\orcid{0000-0002-9336-7551}\inst{\ref{aff18}}
\and A.~D\'iaz-S\'anchez\orcid{0000-0003-0748-4768}\inst{\ref{aff137}}
\and J.~J.~Diaz\orcid{0000-0003-2101-1078}\inst{\ref{aff49}}
\and S.~Di~Domizio\orcid{0000-0003-2863-5895}\inst{\ref{aff34},\ref{aff35}}
\and J.~M.~Diego\orcid{0000-0001-9065-3926}\inst{\ref{aff138}}
\and P.-A.~Duc\orcid{0000-0003-3343-6284}\inst{\ref{aff132}}
\and M.~Y.~Elkhashab\orcid{0000-0001-9306-2603}\inst{\ref{aff7},\ref{aff30},\ref{aff134},\ref{aff8}}
\and A.~Enia\orcid{0000-0002-0200-2857}\inst{\ref{aff18},\ref{aff32}}
\and Y.~Fang\orcid{0000-0002-0334-6950}\inst{\ref{aff68}}
\and A.~G.~Ferrari\orcid{0009-0005-5266-4110}\inst{\ref{aff33}}
\and A.~Finoguenov\orcid{0000-0002-4606-5403}\inst{\ref{aff78}}
\and A.~Fontana\orcid{0000-0003-3820-2823}\inst{\ref{aff46}}
\and F.~Fontanot\orcid{0000-0003-4744-0188}\inst{\ref{aff7},\ref{aff8}}
\and A.~Franco\orcid{0000-0002-4761-366X}\inst{\ref{aff139},\ref{aff140},\ref{aff141}}
\and K.~Ganga\orcid{0000-0001-8159-8208}\inst{\ref{aff89}}
\and J.~Garc\'ia-Bellido\orcid{0000-0002-9370-8360}\inst{\ref{aff126}}
\and T.~Gasparetto\orcid{0000-0002-7913-4866}\inst{\ref{aff46}}
\and E.~Gaztanaga\orcid{0000-0001-9632-0815}\inst{\ref{aff22},\ref{aff109},\ref{aff142}}
\and F.~Giacomini\orcid{0000-0002-3129-2814}\inst{\ref{aff33}}
\and F.~Gianotti\orcid{0000-0003-4666-119X}\inst{\ref{aff18}}
\and G.~Gozaliasl\orcid{0000-0002-0236-919X}\inst{\ref{aff143},\ref{aff78}}
\and M.~Guidi\orcid{0000-0001-9408-1101}\inst{\ref{aff32},\ref{aff18}}
\and C.~M.~Gutierrez\orcid{0000-0001-7854-783X}\inst{\ref{aff21}}
\and A.~Hall\orcid{0000-0002-3139-8651}\inst{\ref{aff50}}
\and C.~Hern\'andez-Monteagudo\orcid{0000-0001-5471-9166}\inst{\ref{aff144},\ref{aff49}}
\and H.~Hildebrandt\orcid{0000-0002-9814-3338}\inst{\ref{aff145}}
\and J.~Hjorth\orcid{0000-0002-4571-2306}\inst{\ref{aff97}}
\and J.~J.~E.~Kajava\orcid{0000-0002-3010-8333}\inst{\ref{aff146},\ref{aff147}}
\and Y.~Kang\orcid{0009-0000-8588-7250}\inst{\ref{aff27}}
\and V.~Kansal\orcid{0000-0002-4008-6078}\inst{\ref{aff148},\ref{aff149}}
\and D.~Karagiannis\orcid{0000-0002-4927-0816}\inst{\ref{aff114},\ref{aff150}}
\and K.~Kiiveri\inst{\ref{aff76}}
\and J.~Kim\orcid{0000-0003-2776-2761}\inst{\ref{aff16}}
\and C.~C.~Kirkpatrick\inst{\ref{aff76}}
\and S.~Kruk\orcid{0000-0001-8010-8879}\inst{\ref{aff52}}
\and L.~Legrand\orcid{0000-0003-0610-5252}\inst{\ref{aff151},\ref{aff14}}
\and M.~Lembo\orcid{0000-0002-5271-5070}\inst{\ref{aff93},\ref{aff114},\ref{aff115}}
\and F.~Lepori\orcid{0009-0000-5061-7138}\inst{\ref{aff152}}
\and G.~Leroy\orcid{0009-0004-2523-4425}\inst{\ref{aff153},\ref{aff88}}
\and G.~F.~Lesci\orcid{0000-0002-4607-2830}\inst{\ref{aff87},\ref{aff18}}
\and J.~Lesgourgues\orcid{0000-0001-7627-353X}\inst{\ref{aff154}}
\and T.~I.~Liaudat\orcid{0000-0002-9104-314X}\inst{\ref{aff155}}
\and S.~J.~Liu\orcid{0000-0001-7680-2139}\inst{\ref{aff61}}
\and J.~Macias-Perez\orcid{0000-0002-5385-2763}\inst{\ref{aff156}}
\and G.~Maggio\orcid{0000-0003-4020-4836}\inst{\ref{aff7}}
\and M.~Magliocchetti\orcid{0000-0001-9158-4838}\inst{\ref{aff61}}
\and F.~Mannucci\orcid{0000-0002-4803-2381}\inst{\ref{aff157}}
\and C.~J.~A.~P.~Martins\orcid{0000-0002-4886-9261}\inst{\ref{aff158},\ref{aff38}}
\and L.~Maurin\orcid{0000-0002-8406-0857}\inst{\ref{aff25}}
\and M.~Miluzio\inst{\ref{aff52},\ref{aff159}}
\and P.~Monaco\orcid{0000-0003-2083-7564}\inst{\ref{aff134},\ref{aff7},\ref{aff30},\ref{aff8}}
\and C.~Moretti\orcid{0000-0003-3314-8936}\inst{\ref{aff7},\ref{aff8},\ref{aff30},\ref{aff31}}
\and G.~Morgante\inst{\ref{aff18}}
\and K.~Naidoo\orcid{0000-0002-9182-1802}\inst{\ref{aff142},\ref{aff75}}
\and A.~Navarro-Alsina\orcid{0000-0002-3173-2592}\inst{\ref{aff85}}
\and S.~Nesseris\orcid{0000-0002-0567-0324}\inst{\ref{aff126}}
\and D.~Paoletti\orcid{0000-0003-4761-6147}\inst{\ref{aff18},\ref{aff65}}
\and F.~Passalacqua\orcid{0000-0002-8606-4093}\inst{\ref{aff11},\ref{aff62}}
\and K.~Paterson\orcid{0000-0001-8340-3486}\inst{\ref{aff19}}
\and L.~Patrizii\inst{\ref{aff33}}
\and A.~Pisani\orcid{0000-0002-6146-4437}\inst{\ref{aff63}}
\and D.~Potter\orcid{0000-0002-0757-5195}\inst{\ref{aff152}}
\and S.~Quai\orcid{0000-0002-0449-8163}\inst{\ref{aff87},\ref{aff18}}
\and M.~Radovich\orcid{0000-0002-3585-866X}\inst{\ref{aff12}}
\and S.~Sacquegna\orcid{0000-0002-8433-6630}\inst{\ref{aff160}}
\and M.~Sahl\'en\orcid{0000-0003-0973-4804}\inst{\ref{aff161}}
\and E.~Sarpa\orcid{0000-0002-1256-655X}\inst{\ref{aff31},\ref{aff125},\ref{aff30}}
\and A.~Schneider\orcid{0000-0001-7055-8104}\inst{\ref{aff152}}
\and D.~Sciotti\orcid{0009-0008-4519-2620}\inst{\ref{aff46},\ref{aff86}}
\and E.~Sellentin\inst{\ref{aff162},\ref{aff10}}
\and L.~C.~Smith\orcid{0000-0002-3259-2771}\inst{\ref{aff15}}
\and K.~Tanidis\orcid{0000-0001-9843-5130}\inst{\ref{aff16}}
\and C.~Tao\orcid{0000-0001-7961-8177}\inst{\ref{aff63}}
\and G.~Testera\inst{\ref{aff35}}
\and R.~Teyssier\orcid{0000-0001-7689-0933}\inst{\ref{aff163}}
\and S.~Tosi\orcid{0000-0002-7275-9193}\inst{\ref{aff34},\ref{aff35},\ref{aff29}}
\and A.~Troja\orcid{0000-0003-0239-4595}\inst{\ref{aff11},\ref{aff62}}
\and M.~Tucci\inst{\ref{aff27}}
\and A.~Venhola\orcid{0000-0001-6071-4564}\inst{\ref{aff164}}
\and D.~Vergani\orcid{0000-0003-0898-2216}\inst{\ref{aff18}}
\and G.~Verza\orcid{0000-0002-1886-8348}\inst{\ref{aff165}}
\and P.~Vielzeuf\orcid{0000-0003-2035-9339}\inst{\ref{aff63}}
\and N.~A.~Walton\orcid{0000-0003-3983-8778}\inst{\ref{aff15}}}
										   
%%%% please do not edit the affiliation list -- contact ECEB Bureau for changes
\institute{Kapteyn Astronomical Institute, University of Groningen, PO Box 800, 9700 AV Groningen, The Netherlands\label{aff1}
\and
Cosmic Dawn Center (DAWN)\label{aff2}
\and
Cosmic Dawn Center (DAWN), Denmark\label{aff3}
\and
Niels Bohr Institute, University of Copenhagen, Jagtvej 128, 2200 Copenhagen, Denmark\label{aff4}
\and
MIT Kavli Institute for Astrophysics and Space Research, Massachusetts Institute of Technology, Cambridge, MA 02139, USA\label{aff5}
\and
Institute for Astronomy, University of Hawaii, 2680 Woodlawn Drive, Honolulu, HI 96822, USA\label{aff6}
\and
INAF-Osservatorio Astronomico di Trieste, Via G. B. Tiepolo 11, 34143 Trieste, Italy\label{aff7}
\and
IFPU, Institute for Fundamental Physics of the Universe, via Beirut 2, 34151 Trieste, Italy\label{aff8}
\and
Jodrell Bank Centre for Astrophysics, Department of Physics and Astronomy, University of Manchester, Oxford Road, Manchester M13 9PL, UK\label{aff9}
\and
Leiden Observatory, Leiden University, Einsteinweg 55, 2333 CC Leiden, The Netherlands\label{aff10}
\and
Dipartimento di Fisica e Astronomia "G. Galilei", Universit\`a di Padova, Via Marzolo 8, 35131 Padova, Italy\label{aff11}
\and
INAF-Osservatorio Astronomico di Padova, Via dell'Osservatorio 5, 35122 Padova, Italy\label{aff12}
\and
Aix-Marseille Universit\'e, CNRS, CNES, LAM, Marseille, France\label{aff13}
\and
Kavli Institute for Cosmology Cambridge, Madingley Road, Cambridge, CB3 0HA, UK\label{aff14}
\and
Institute of Astronomy, University of Cambridge, Madingley Road, Cambridge CB3 0HA, UK\label{aff15}
\and
Department of Physics, Oxford University, Keble Road, Oxford OX1 3RH, UK\label{aff16}
\and
Department of Physics \& Astronomy, University of Sussex, Brighton BN1 9QH, UK\label{aff17}
\and
INAF-Osservatorio di Astrofisica e Scienza dello Spazio di Bologna, Via Piero Gobetti 93/3, 40129 Bologna, Italy\label{aff18}
\and
Max-Planck-Institut f\"ur Astronomie, K\"onigstuhl 17, 69117 Heidelberg, Germany\label{aff19}
\and
Department of Physics and Astronomy, University of British Columbia, Vancouver, BC V6T 1Z1, Canada\label{aff20}
\and
 Instituto de Astrof\'{\i}sica de Canarias, E-38205 La Laguna; Universidad de La Laguna, Dpto. Astrof\'\i sica, E-38206 La Laguna, Tenerife, Spain\label{aff21}
\and
Institute of Space Sciences (ICE, CSIC), Campus UAB, Carrer de Can Magrans, s/n, 08193 Barcelona, Spain\label{aff22}
\and
School of Physics \& Astronomy, University of Southampton, Highfield Campus, Southampton SO17 1BJ, UK\label{aff23}
\and
Univ. Lille, CNRS, Centrale Lille, UMR 9189 CRIStAL, 59000 Lille, France\label{aff24}
\and
Universit\'e Paris-Saclay, CNRS, Institut d'astrophysique spatiale, 91405, Orsay, France\label{aff25}
\and
Institute for Particle Physics and Astrophysics, Dept. of Physics, ETH Zurich, Wolfgang-Pauli-Strasse 27, 8093 Zurich, Switzerland\label{aff26}
\and
Department of Astronomy, University of Geneva, ch. d'Ecogia 16, 1290 Versoix, Switzerland\label{aff27}
\and
School of Mathematics and Physics, University of Surrey, Guildford, Surrey, GU2 7XH, UK\label{aff28}
\and
INAF-Osservatorio Astronomico di Brera, Via Brera 28, 20122 Milano, Italy\label{aff29}
\and
INFN, Sezione di Trieste, Via Valerio 2, 34127 Trieste TS, Italy\label{aff30}
\and
SISSA, International School for Advanced Studies, Via Bonomea 265, 34136 Trieste TS, Italy\label{aff31}
\and
Dipartimento di Fisica e Astronomia, Universit\`a di Bologna, Via Gobetti 93/2, 40129 Bologna, Italy\label{aff32}
\and
INFN-Sezione di Bologna, Viale Berti Pichat 6/2, 40127 Bologna, Italy\label{aff33}
\and
Dipartimento di Fisica, Universit\`a di Genova, Via Dodecaneso 33, 16146, Genova, Italy\label{aff34}
\and
INFN-Sezione di Genova, Via Dodecaneso 33, 16146, Genova, Italy\label{aff35}
\and
Department of Physics "E. Pancini", University Federico II, Via Cinthia 6, 80126, Napoli, Italy\label{aff36}
\and
INAF-Osservatorio Astronomico di Capodimonte, Via Moiariello 16, 80131 Napoli, Italy\label{aff37}
\and
Instituto de Astrof\'isica e Ci\^encias do Espa\c{c}o, Universidade do Porto, CAUP, Rua das Estrelas, PT4150-762 Porto, Portugal\label{aff38}
\and
Faculdade de Ci\^encias da Universidade do Porto, Rua do Campo de Alegre, 4150-007 Porto, Portugal\label{aff39}
\and
European Southern Observatory, Karl-Schwarzschild-Str.~2, 85748 Garching, Germany\label{aff40}
\and
European Space Agency/ESTEC, Keplerlaan 1, 2201 AZ Noordwijk, The Netherlands\label{aff41}
\and
INAF-Osservatorio Astrofisico di Torino, Via Osservatorio 20, 10025 Pino Torinese (TO), Italy\label{aff42}
\and
INAF-IASF Milano, Via Alfonso Corti 12, 20133 Milano, Italy\label{aff43}
\and
Centro de Investigaciones Energ\'eticas, Medioambientales y Tecnol\'ogicas (CIEMAT), Avenida Complutense 40, 28040 Madrid, Spain\label{aff44}
\and
Port d'Informaci\'{o} Cient\'{i}fica, Campus UAB, C. Albareda s/n, 08193 Bellaterra (Barcelona), Spain\label{aff45}
\and
INAF-Osservatorio Astronomico di Roma, Via Frascati 33, 00078 Monteporzio Catone, Italy\label{aff46}
\and
INFN section of Naples, Via Cinthia 6, 80126, Napoli, Italy\label{aff47}
\and
Dipartimento di Fisica e Astronomia "Augusto Righi" - Alma Mater Studiorum Universit\`a di Bologna, Viale Berti Pichat 6/2, 40127 Bologna, Italy\label{aff48}
\and
Instituto de Astrof\'{\i}sica de Canarias, E-38205 La Laguna, Tenerife, Spain\label{aff49}
\and
Institute for Astronomy, University of Edinburgh, Royal Observatory, Blackford Hill, Edinburgh EH9 3HJ, UK\label{aff50}
\and
European Space Agency/ESRIN, Largo Galileo Galilei 1, 00044 Frascati, Roma, Italy\label{aff51}
\and
ESAC/ESA, Camino Bajo del Castillo, s/n., Urb. Villafranca del Castillo, 28692 Villanueva de la Ca\~nada, Madrid, Spain\label{aff52}
\and
Universit\'e Claude Bernard Lyon 1, CNRS/IN2P3, IP2I Lyon, UMR 5822, Villeurbanne, F-69100, France\label{aff53}
\and
Institut de Ci\`{e}ncies del Cosmos (ICCUB), Universitat de Barcelona (IEEC-UB), Mart\'{i} i Franqu\`{e}s 1, 08028 Barcelona, Spain\label{aff54}
\and
Instituci\'o Catalana de Recerca i Estudis Avan\c{c}ats (ICREA), Passeig de Llu\'{\i}s Companys 23, 08010 Barcelona, Spain\label{aff55}
\and
Institut de Ciencies de l'Espai (IEEC-CSIC), Campus UAB, Carrer de Can Magrans, s/n Cerdanyola del Vall\'es, 08193 Barcelona, Spain\label{aff56}
\and
UCB Lyon 1, CNRS/IN2P3, IUF, IP2I Lyon, 4 rue Enrico Fermi, 69622 Villeurbanne, France\label{aff57}
\and
Mullard Space Science Laboratory, University College London, Holmbury St Mary, Dorking, Surrey RH5 6NT, UK\label{aff58}
\and
Departamento de F\'isica, Faculdade de Ci\^encias, Universidade de Lisboa, Edif\'icio C8, Campo Grande, PT1749-016 Lisboa, Portugal\label{aff59}
\and
Instituto de Astrof\'isica e Ci\^encias do Espa\c{c}o, Faculdade de Ci\^encias, Universidade de Lisboa, Campo Grande, 1749-016 Lisboa, Portugal\label{aff60}
\and
INAF-Istituto di Astrofisica e Planetologia Spaziali, via del Fosso del Cavaliere, 100, 00100 Roma, Italy\label{aff61}
\and
INFN-Padova, Via Marzolo 8, 35131 Padova, Italy\label{aff62}
\and
Aix-Marseille Universit\'e, CNRS/IN2P3, CPPM, Marseille, France\label{aff63}
\and
Space Science Data Center, Italian Space Agency, via del Politecnico snc, 00133 Roma, Italy\label{aff64}
\and
INFN-Bologna, Via Irnerio 46, 40126 Bologna, Italy\label{aff65}
\and
University Observatory, LMU Faculty of Physics, Scheinerstr.~1, 81679 Munich, Germany\label{aff66}
\and
Max Planck Institute for Extraterrestrial Physics, Giessenbachstr. 1, 85748 Garching, Germany\label{aff67}
\and
Universit\"ats-Sternwarte M\"unchen, Fakult\"at f\"ur Physik, Ludwig-Maximilians-Universit\"at M\"unchen, Scheinerstr.~1, 81679 M\"unchen, Germany\label{aff68}
\and
Institute of Theoretical Astrophysics, University of Oslo, P.O. Box 1029 Blindern, 0315 Oslo, Norway\label{aff69}
\and
Jet Propulsion Laboratory, California Institute of Technology, 4800 Oak Grove Drive, Pasadena, CA, 91109, USA\label{aff70}
\and
Department of Physics, Lancaster University, Lancaster, LA1 4YB, UK\label{aff71}
\and
Felix Hormuth Engineering, Goethestr. 17, 69181 Leimen, Germany\label{aff72}
\and
Technical University of Denmark, Elektrovej 327, 2800 Kgs. Lyngby, Denmark\label{aff73}
\and
NASA Goddard Space Flight Center, Greenbelt, MD 20771, USA\label{aff74}
\and
Department of Physics and Astronomy, University College London, Gower Street, London WC1E 6BT, UK\label{aff75}
\and
Department of Physics and Helsinki Institute of Physics, Gustaf H\"allstr\"omin katu 2, University of Helsinki, 00014 Helsinki, Finland\label{aff76}
\and
Universit\'e de Gen\`eve, D\'epartement de Physique Th\'eorique and Centre for Astroparticle Physics, 24 quai Ernest-Ansermet, CH-1211 Gen\`eve 4, Switzerland\label{aff77}
\and
Department of Physics, P.O. Box 64, University of Helsinki, 00014 Helsinki, Finland\label{aff78}
\and
Helsinki Institute of Physics, Gustaf H{\"a}llstr{\"o}min katu 2, University of Helsinki, 00014 Helsinki, Finland\label{aff79}
\and
Laboratoire d'etude de l'Univers et des phenomenes eXtremes, Observatoire de Paris, Universit\'e PSL, Sorbonne Universit\'e, CNRS, 92190 Meudon, France\label{aff80}
\and
SKAO, Jodrell Bank, Lower Withington, Macclesfield SK11 9FT, UK\label{aff81}
\and
Centre de Calcul de l'IN2P3/CNRS, 21 avenue Pierre de Coubertin 69627 Villeurbanne Cedex, France\label{aff82}
\and
Dipartimento di Fisica "Aldo Pontremoli", Universit\`a degli Studi di Milano, Via Celoria 16, 20133 Milano, Italy\label{aff83}
\and
INFN-Sezione di Milano, Via Celoria 16, 20133 Milano, Italy\label{aff84}
\and
Universit\"at Bonn, Argelander-Institut f\"ur Astronomie, Auf dem H\"ugel 71, 53121 Bonn, Germany\label{aff85}
\and
INFN-Sezione di Roma, Piazzale Aldo Moro, 2 - c/o Dipartimento di Fisica, Edificio G. Marconi, 00185 Roma, Italy\label{aff86}
\and
Dipartimento di Fisica e Astronomia "Augusto Righi" - Alma Mater Studiorum Universit\`a di Bologna, via Piero Gobetti 93/2, 40129 Bologna, Italy\label{aff87}
\and
Department of Physics, Institute for Computational Cosmology, Durham University, South Road, Durham, DH1 3LE, UK\label{aff88}
\and
Universit\'e Paris Cit\'e, CNRS, Astroparticule et Cosmologie, 75013 Paris, France\label{aff89}
\and
CNRS-UCB International Research Laboratory, Centre Pierre Bin\'etruy, IRL2007, CPB-IN2P3, Berkeley, USA\label{aff90}
\and
University of Applied Sciences and Arts of Northwestern Switzerland, School of Engineering, 5210 Windisch, Switzerland\label{aff91}
\and
Institut d'Astrophysique de Paris, 98bis Boulevard Arago, 75014, Paris, France\label{aff92}
\and
Institut d'Astrophysique de Paris, UMR 7095, CNRS, and Sorbonne Universit\'e, 98 bis boulevard Arago, 75014 Paris, France\label{aff93}
\and
Institute of Physics, Laboratory of Astrophysics, Ecole Polytechnique F\'ed\'erale de Lausanne (EPFL), Observatoire de Sauverny, 1290 Versoix, Switzerland\label{aff94}
\and
Telespazio UK S.L. for European Space Agency (ESA), Camino bajo del Castillo, s/n, Urbanizacion Villafranca del Castillo, Villanueva de la Ca\~nada, 28692 Madrid, Spain\label{aff95}
\and
Institut de F\'{i}sica d'Altes Energies (IFAE), The Barcelona Institute of Science and Technology, Campus UAB, 08193 Bellaterra (Barcelona), Spain\label{aff96}
\and
DARK, Niels Bohr Institute, University of Copenhagen, Jagtvej 155, 2200 Copenhagen, Denmark\label{aff97}
\and
Waterloo Centre for Astrophysics, University of Waterloo, Waterloo, Ontario N2L 3G1, Canada\label{aff98}
\and
Department of Physics and Astronomy, University of Waterloo, Waterloo, Ontario N2L 3G1, Canada\label{aff99}
\and
Perimeter Institute for Theoretical Physics, Waterloo, Ontario N2L 2Y5, Canada\label{aff100}
\and
Universit\'e Paris-Saclay, Universit\'e Paris Cit\'e, CEA, CNRS, AIM, 91191, Gif-sur-Yvette, France\label{aff101}
\and
Centre National d'Etudes Spatiales -- Centre spatial de Toulouse, 18 avenue Edouard Belin, 31401 Toulouse Cedex 9, France\label{aff102}
\and
Institute of Space Science, Str. Atomistilor, nr. 409 M\u{a}gurele, Ilfov, 077125, Romania\label{aff103}
\and
Institut f\"ur Theoretische Physik, University of Heidelberg, Philosophenweg 16, 69120 Heidelberg, Germany\label{aff104}
\and
Institut de Recherche en Astrophysique et Plan\'etologie (IRAP), Universit\'e de Toulouse, CNRS, UPS, CNES, 14 Av. Edouard Belin, 31400 Toulouse, France\label{aff105}
\and
Universit\'e St Joseph; Faculty of Sciences, Beirut, Lebanon\label{aff106}
\and
Departamento de F\'isica, FCFM, Universidad de Chile, Blanco Encalada 2008, Santiago, Chile\label{aff107}
\and
Universit\"at Innsbruck, Institut f\"ur Astro- und Teilchenphysik, Technikerstr. 25/8, 6020 Innsbruck, Austria\label{aff108}
\and
Institut d'Estudis Espacials de Catalunya (IEEC),  Edifici RDIT, Campus UPC, 08860 Castelldefels, Barcelona, Spain\label{aff109}
\and
Satlantis, University Science Park, Sede Bld 48940, Leioa-Bilbao, Spain\label{aff110}
\and
Infrared Processing and Analysis Center, California Institute of Technology, Pasadena, CA 91125, USA\label{aff111}
\and
Instituto de Astrof\'isica e Ci\^encias do Espa\c{c}o, Faculdade de Ci\^encias, Universidade de Lisboa, Tapada da Ajuda, 1349-018 Lisboa, Portugal\label{aff112}
\and
Universidad Polit\'ecnica de Cartagena, Departamento de Electr\'onica y Tecnolog\'ia de Computadoras,  Plaza del Hospital 1, 30202 Cartagena, Spain\label{aff113}
\and
Dipartimento di Fisica e Scienze della Terra, Universit\`a degli Studi di Ferrara, Via Giuseppe Saragat 1, 44122 Ferrara, Italy\label{aff114}
\and
Istituto Nazionale di Fisica Nucleare, Sezione di Ferrara, Via Giuseppe Saragat 1, 44122 Ferrara, Italy\label{aff115}
\and
INAF, Istituto di Radioastronomia, Via Piero Gobetti 101, 40129 Bologna, Italy\label{aff116}
\and
Astronomical Observatory of the Autonomous Region of the Aosta Valley (OAVdA), Loc. Lignan 39, I-11020, Nus (Aosta Valley), Italy\label{aff117}
\and
Universit\'e C\^{o}te d'Azur, Observatoire de la C\^{o}te d'Azur, CNRS, Laboratoire Lagrange, Bd de l'Observatoire, CS 34229, 06304 Nice cedex 4, France\label{aff118}
\and
Universit\'e PSL, Observatoire de Paris, Sorbonne Universit\'e, CNRS, LERMA, 75014, Paris, France\label{aff119}
\and
Universit\'e Paris-Cit\'e, 5 Rue Thomas Mann, 75013, Paris, France\label{aff120}
\and
Dipartimento di Fisica, Sapienza Universit\`a di Roma, Piazzale Aldo Moro 2, 00185 Roma, Italy\label{aff121}
\and
Aurora Technology for European Space Agency (ESA), Camino bajo del Castillo, s/n, Urbanizacion Villafranca del Castillo, Villanueva de la Ca\~nada, 28692 Madrid, Spain\label{aff122}
\and
Zentrum f\"ur Astronomie, Universit\"at Heidelberg, Philosophenweg 12, 69120 Heidelberg, Germany\label{aff123}
\and
ICL, Junia, Universit\'e Catholique de Lille, LITL, 59000 Lille, France\label{aff124}
\and
ICSC - Centro Nazionale di Ricerca in High Performance Computing, Big Data e Quantum Computing, Via Magnanelli 2, Bologna, Italy\label{aff125}
\and
Instituto de F\'isica Te\'orica UAM-CSIC, Campus de Cantoblanco, 28049 Madrid, Spain\label{aff126}
\and
CERCA/ISO, Department of Physics, Case Western Reserve University, 10900 Euclid Avenue, Cleveland, OH 44106, USA\label{aff127}
\and
Technical University of Munich, TUM School of Natural Sciences, Physics Department, James-Franck-Str.~1, 85748 Garching, Germany\label{aff128}
\and
Max-Planck-Institut f\"ur Astrophysik, Karl-Schwarzschild-Str.~1, 85748 Garching, Germany\label{aff129}
\and
Laboratoire Univers et Th\'eorie, Observatoire de Paris, Universit\'e PSL, Universit\'e Paris Cit\'e, CNRS, 92190 Meudon, France\label{aff130}
\and
Departamento de F{\'\i}sica Fundamental. Universidad de Salamanca. Plaza de la Merced s/n. 37008 Salamanca, Spain\label{aff131}
\and
Universit\'e de Strasbourg, CNRS, Observatoire astronomique de Strasbourg, UMR 7550, 67000 Strasbourg, France\label{aff132}
\and
Center for Data-Driven Discovery, Kavli IPMU (WPI), UTIAS, The University of Tokyo, Kashiwa, Chiba 277-8583, Japan\label{aff133}
\and
Dipartimento di Fisica - Sezione di Astronomia, Universit\`a di Trieste, Via Tiepolo 11, 34131 Trieste, Italy\label{aff134}
\and
California Institute of Technology, 1200 E California Blvd, Pasadena, CA 91125, USA\label{aff135}
\and
Department of Physics \& Astronomy, University of California Irvine, Irvine CA 92697, USA\label{aff136}
\and
Departamento F\'isica Aplicada, Universidad Polit\'ecnica de Cartagena, Campus Muralla del Mar, 30202 Cartagena, Murcia, Spain\label{aff137}
\and
Instituto de F\'isica de Cantabria, Edificio Juan Jord\'a, Avenida de los Castros, 39005 Santander, Spain\label{aff138}
\and
INFN, Sezione di Lecce, Via per Arnesano, CP-193, 73100, Lecce, Italy\label{aff139}
\and
Department of Mathematics and Physics E. De Giorgi, University of Salento, Via per Arnesano, CP-I93, 73100, Lecce, Italy\label{aff140}
\and
INAF-Sezione di Lecce, c/o Dipartimento Matematica e Fisica, Via per Arnesano, 73100, Lecce, Italy\label{aff141}
\and
Institute of Cosmology and Gravitation, University of Portsmouth, Portsmouth PO1 3FX, UK\label{aff142}
\and
Department of Computer Science, Aalto University, PO Box 15400, Espoo, FI-00 076, Finland\label{aff143}
\and
Universidad de La Laguna, Dpto. Astrof\'\i sica, E-38206 La Laguna, Tenerife, Spain\label{aff144}
\and
Ruhr University Bochum, Faculty of Physics and Astronomy, Astronomical Institute (AIRUB), German Centre for Cosmological Lensing (GCCL), 44780 Bochum, Germany\label{aff145}
\and
Department of Physics and Astronomy, Vesilinnantie 5, University of Turku, 20014 Turku, Finland\label{aff146}
\and
Serco for European Space Agency (ESA), Camino bajo del Castillo, s/n, Urbanizacion Villafranca del Castillo, Villanueva de la Ca\~nada, 28692 Madrid, Spain\label{aff147}
\and
ARC Centre of Excellence for Dark Matter Particle Physics, Melbourne, Australia\label{aff148}
\and
Centre for Astrophysics \& Supercomputing, Swinburne University of Technology,  Hawthorn, Victoria 3122, Australia\label{aff149}
\and
Department of Physics and Astronomy, University of the Western Cape, Bellville, Cape Town, 7535, South Africa\label{aff150}
\and
DAMTP, Centre for Mathematical Sciences, Wilberforce Road, Cambridge CB3 0WA, UK\label{aff151}
\and
Department of Astrophysics, University of Zurich, Winterthurerstrasse 190, 8057 Zurich, Switzerland\label{aff152}
\and
Department of Physics, Centre for Extragalactic Astronomy, Durham University, South Road, Durham, DH1 3LE, UK\label{aff153}
\and
Institute for Theoretical Particle Physics and Cosmology (TTK), RWTH Aachen University, 52056 Aachen, Germany\label{aff154}
\and
IRFU, CEA, Universit\'e Paris-Saclay 91191 Gif-sur-Yvette Cedex, France\label{aff155}
\and
Univ. Grenoble Alpes, CNRS, Grenoble INP, LPSC-IN2P3, 53, Avenue des Martyrs, 38000, Grenoble, France\label{aff156}
\and
INAF-Osservatorio Astrofisico di Arcetri, Largo E. Fermi 5, 50125, Firenze, Italy\label{aff157}
\and
Centro de Astrof\'{\i}sica da Universidade do Porto, Rua das Estrelas, 4150-762 Porto, Portugal\label{aff158}
\and
HE Space for European Space Agency (ESA), Camino bajo del Castillo, s/n, Urbanizacion Villafranca del Castillo, Villanueva de la Ca\~nada, 28692 Madrid, Spain\label{aff159}
\and
INAF - Osservatorio Astronomico d'Abruzzo, Via Maggini, 64100, Teramo, Italy\label{aff160}
\and
Theoretical astrophysics, Department of Physics and Astronomy, Uppsala University, Box 516, 751 37 Uppsala, Sweden\label{aff161}
\and
Mathematical Institute, University of Leiden, Einsteinweg 55, 2333 CA Leiden, The Netherlands\label{aff162}
\and
Department of Astrophysical Sciences, Peyton Hall, Princeton University, Princeton, NJ 08544, USA\label{aff163}
\and
Space physics and astronomy research unit, University of Oulu, Pentti Kaiteran katu 1, FI-90014 Oulu, Finland\label{aff164}
\and
Center for Computational Astrophysics, Flatiron Institute, 162 5th Avenue, 10010, New York, NY, USA\label{aff165}}    

 % \date{}

% 
% Put your abstract here
%
\abstract{
Probing the presence and properties of massive galaxies at high redshift is one of the most critical tests for galaxy formation models. In this work, we search for galaxies with stellar masses \( M_* > 10^{10.25} \, \rm{M_\odot} \) at \( z \in [5,7]\), i.e., towards the end of the Epoch of Reionisation, over a total of \( \sim 23\, \mathrm{deg}^2\) in two of the Euclid Quick Data Release (Q1) fields: the Euclid Deep Field North and Fornax (EDF-N and EDF-F). In addition to the \textit{Euclid} photometry, we incorporate \textit{Spitzer} Infrared Camera (IRAC) and ground-based optical data to perform spectral energy distribution (SED) fitting, obtaining photometric redshifts and derived physical parameters. After applying rigorous selection criteria,  we identify a conservative sample of 145 candidate massive galaxies with \( M_* > 10^{10.25} \, \rm{M_\odot} \) at \( z \in [5,7] \), including 5 objects with \( M_* > 10^{11} \, \rm{M_\odot} \).  This makes for a surface density of about 6.3~deg$^{-2}$ at $z\in [5,7]$, which should be considered a lower limit because of the current depth of the \textit{Euclid} data ($\HE<24, \, 5\sigma$ in Q1). We find that the inferred stellar masses are consistent with galaxy formation models with standard star-formation efficiencies. These massive galaxies have colour excess $E(B-V)$ values up to 0.75, indicating significant dust attenuation in some of them. In addition, half of the massive galaxies have best-fit ages comparable to the age of the Universe at those redshifts, which suggests that their progenitors were formed very early in cosmic time. About 78\%  of the massive galaxies lie on the star-forming main sequence (MS) in the SFR–\( M_* \) plane, $\sim$12\%  are found in the starburst region, and 10\% in the transition zone between the MS and starbursts. We find no significant evidence for outshining or AGN contamination that could account for the elevated specific star-formation rates (sSFR) observed in the \(\sim12\%\) of galaxies classified as starbursts.
}

%
% Provide up to five key words:
%
%
% Add short versions of title and author list for page headings
%
    \titlerunning{Euclid Q1: Massive galaxies at the
end of the EoR}
   \authorrunning{Euclid Collaboration: Navarro-Carrera et al.}
   
   \maketitle
%
%-------------------------------------------------------------------
%
%
%   Start the main text of your paper here
%
\nolinenumbers
\section{\label{sc:Intro}Introduction}

 Understanding when and how the most massive galaxies in the Universe formed is one of the most important goals in extragalactic astronomy. At low redshifts, most massive galaxies contain old stellar populations, indicating early formation times \citep[e.g.,][]{franx_significant_2003, cimatti_old_2004, fontana_k20_2004, saracco_density_2005, pozzetti_vimos_2007, ilbert_galaxy_2010}. Moreover, a significant fraction of them were not only present, but already massive at $z\gsim3$ \citep[e.g.,][]{caputi_stellar_2011, davidzon_cosmos2015_2017, marsan_number_2022}, indicating that stellar mass assembly in these sources must have happened quite rapidly in the first few billion years of cosmic time.

These observational results constitute important constraints for galaxy formation models, which need to invoke efficient star-formation processes and to explain the formation of the most massive galaxies \citep{di_matteo_energy_2005, croton_evolution_2006} at such early times. Together with these, they need to incorporate feedback from active galactic nuclei (AGN) to get a full understanding of their evolution, as these mechanisms are known to be at play since at least $z\sim 4$ \citep[e.g.,][]{saxena_widespread_2024}.

Therefore, pushing the search of these most massive galaxies to the very first billion years, towards the Epoch of Reionisation (EoR),  is essential to understand when the formation of these objects first happened in the early Universe. 

Observations conducted over the past decade indicated that galaxies with $M_* > 10^{11} \, \rm M_\odot$ were very rare at $z \gsim 5.5$ \citep[e.g.,][]{caputi_spitzer_2015, weaver_cosmos2020_2022}. Until now,  galaxy surveys deep enough to search for such galaxies have mostly been limited to relatively small areas of the sky ($\sim$~1~deg$^2$), restricting the possibility of identifying significant samples of such massive galaxies at high redshifts. In somewhat shallower wider-area surveys, a small number of massive galaxies have been identified at $z\sim 7$ \citep{Varadaraj25}, but none with $M_* > 10^{11} \, \rm M_\odot$. A systematic search for these objects requires deep near-/mid-infrared (IR) imaging over larger areas of the sky. Indeed, including up to mid-IR wavelengths ($\sim 5 \, \rm \mu m$) in the wide-area imaging is essential for a proper stellar-mass determination at high redshifts. This is because the light of old stars, which is mainly emitted in the rest-frame optical, reaches us shifted into the mid-IR.

The \textit{Euclid} Space Telescope \citep{EuclidSkyOverview} is now providing us with the first deep and wide-area near-IR images from which we can select massive galaxy candidates at high $z$. These images are part of the quick (Q1) data release of the Euclid Deep Survey, which will eventually cover $\sim 53 \, \rm deg^2$ of the sky, over three fields, namely the Euclid Deep Field North (EDF-N), South (EDF-S), and Fornax (EDF-F).  Here we identify galaxies with $M_* > 10^{10.25} \, \rm M_\odot$, restricting our analysis to a total of $\sim 23 \, \rm deg^2$ within EDF-N and EDF-F, because this is the area with ancillary mid-IR coverage from the Spitzer Space Telescope Infrared Array Camera \citep[IRAC;][]{fazio_infrared_2004} and deep optical data from the Hawaii Twenty Square Degree survey \citep[H2O,][]{EP-McPartland}.

The layout of this paper is as follows. In Sect.~2, we present the datasets and explain our photometric catalogue construction. In Sect.~3, we give details of our spectral energy distribution (SED) fitting, which yields the estimation of photometric redshifts and stellar masses. Our massive galaxy candidates at $z\in[5,7]$ are selected based on this output, as we explain in Sect.~4, where we also present their basic properties. In Sect.~5, we analyse our candidates within the context of the star-formation versus stellar mass (SFR-$M_*$) plane. Finally, in Sect.~6 we discuss our findings and present our general conclusions. We adopt throughout a standard cosmology with $\Omega_{\rm m}=0.3$, $\Omega_\Lambda=0.7$, and $H_0=70 \, \rm km \, s^{-1} \, Mpc^{-1}$. All magnitudes are given in the AB system \citep{oke_secondary_1983}.

%-------------------------------------------------------------------

\section{\label{sc:data}Datasets and photometric catalogue construction}

For this work, we adopt the DAWN \textit{Euclid} images and catalogues corresponding to Q1 \citep{Q1cite, EP-Aussel,Q1-TP002,Q1-TP003, Q1-SP047} in the EDF-N and EDF-F. These catalogues contain \textit{Euclid} photometry in three near-IR (NISP; \citealp{EuclidSkyNISP}) bands, namely $\YE$, $\JE$,  and $\HE$, as well as optical photometry in a single ($\IE$) filter (VIS instrument; \citealp{EuclidSkyVIS}).

In addition, ancillary imaging data are available from the H20 Survey (CFHT $u$, HSC $grizy$; \citealp{EP-Zalesky}), and the Spitzer Legacy Survey (3.6 and 4.5~$\mu$m; \citealp{moneti_euclid_2022}). The list of all considered pass bands, along with their effective wavelengths and depths, are listed in Table \ref{tab:depth}. We consider the full EDF-N and EDF-F area  with IRAC coverage for our analysis, for a total of $\sim 23 \rm \, deg^2$.

\begin{table}[t]
    \caption{Imaging depths and effective wavelengths \citep[obtained from][]{Schirmer-EP18} for all filters used in our study. These depths were calculated in apertures of size two times the FWHM. Ancillary data (IRAC, HSC, and CFHT) are quoted for 2$\arcsec$ apertures \citep{EP-McPartland} in the full depth area, which encompasses to $\sim 75 \%$ of our sample. When different, depths are reported for EDF-N first and EDF-F second. Note there is no HSC $y$ coverage for the EDF-F.}
    \renewcommand{\arraystretch}{1.1}
    \label{tab:depth}
    \centering
    \begin{tabular}{lccc}
        \hline \hline
        Band & $\lambda_{\text{eff}}$ [\AA] & $5\sigma$ depth [AB] \\
        \hline
        CFHT $u$   & 3720    & 26.7/26.4 \\
        HSC  $g$   & 4759    & 27.2/27.2 \\
        HSC  $r$   & 6166    & 27.4 \\
        HSC  $i$   & 7682    & 26.8/27.0 \\
        HSC  $z$   & 8906    & 26.2/25.1 \\
        HSC  $y$   & 9790    & 24.5/- \\
        VIS  $\IE$ & 7150    & 25.4 \\
        NISP $\YE$ & 10\,809 & 24.0  \\
        NISP $\JE$ & 13\,673 & 24.0  \\
        NISP $\HE$ & 17\,714 & 24.0  \\
        IRAC1 & 35\,500 & 24.9/25.1  \\
        IRAC2 & 44\,930 & 24.8/24.9 \\
        \hline
    \end{tabular}
\end{table}

\subsection{\label{ssc:catalog} Source detection and photometric catalogue}

To construct our parent source catalogue, we include all \textit{Euclid} NISP-detected sources in the EDF-N and EDF-F fields. Our starting point is the Q1 DAWN catalogue\footnote{The DAWN Q1 catalogue can be found in the repository \url{https://dawn.calet.org/q1/}.}, based on detections in a NISP 3-filter stack and limited to the area with deep \textit{Spitzer}/IRAC coverage \citep{moneti_euclid_2022}. 

The \textit{Euclid} data used here include only the Q1 observation set. NISP mosaics were drizzled from individual exposures using producing images with $\ang{; ;0.2}$ pixel scale. Source detection, deblending and background subtraction were carried out using \textsc{SEP} \citep{barbary_sep_2016}, more details can be found in appendix \ref{sec_app:detection}.

We extracted photometry for all sources using \textsc{The Farmer} \citep{weaver_farmer_2023}. Briefly, NISP images were used to construct surface brightness profiles for each source, accounting for the contribution of neighbouring objects. Using spatially variable PSF models extracted from the mosaics, photometry in the remaining bands was measured using the \textit{Euclid}-based priors. This approach yields optimal photometry for the low-resolution ground-based optical data, and is particularly beneficial for the \textit{Spitzer}/IRAC images.
 
The resulting source counts are 5\,394\,301 in EDF-N, and 3\,733\,178 in EDF-F.

\subsection{\label{ssc:irac_datalog} IRAC source catalogue}

We constructed a complementary source catalogue based on detections in IRAC. For these sources we measured the IRAC photometry from the \textit{Spitzer}/IRAC [ch1] (IRAC1) and [ch2] (IRAC2) images used in the DAWN catalogue, using \textsc{SExtractor} \citep{bertin_sextractor_1996} following an identical procedure to the one detailed in \cite{Moneti-EP17}.

Briefly, we first constructed an inverse-variance weighted detection image comprising  the two considered IRAC channels. We then ran \textsc{SExtractor} in dual mode to extract the aperture photometry of sources in both bands. For sources brighter than $23$~AB~mag, we did not find any significant departure with respect to the \textsc{Farmer} (\citealp{weaver_farmer_2023}) photometry.

The resulting IRAC source counts are 1\,552\,985 in EDF-N, and 1\,348\,399 in EDF-F. We do not employ our IRAC photometry and solely use the information on the detected sources to aid the selection of non-spurious and non-blended sources \citep{weaver_euclid_2025}.

\begin{table}[t]
    \centering
    \caption{Parameter values used for \textsc{LePHARE} and \textsc{BAGPIPES} SED fitting codes.}
    \label{tab:sedfit_params}
    \renewcommand{\arraystretch}{1.2}
    \setlength{\tabcolsep}{2.2mm}
    \begin{tabular}{l|cc}
        \hline\hline
        \textbf{Parameter} & \textbf{\textsc{LePHARE}} & \textbf{\textsc{BAGPIPES}} \\
        \hline
        Templates & \multicolumn{2}{c}{\textsc{BC03}}  \\ 
        SFH & \multicolumn{2}{c}{exponential}  \\ 
        e-folding time (Gyr) & \multicolumn{2}{c}{[0.0001,15]}  \\
        \hline
        Redshift & {[0,12]} & {[0,12]}  \\
        $\logten({M_*}/\rm{M_{\odot}})$& -- & [1,13]  \\
        Metallicity ($\rm Z_{\odot}$) & [0.2, 1] & [0.001,1]  \\
        Age (Gyr) & {[0.01, 15]} & {[0.01,15]}   \\
        $\logten U$ & --  & $[-4, -0.5]$   \\
        \hline
        Extinction law & \multicolumn{2}{c}{\citet{calzetti_dust_2000}}   \\
        $A_V$ (AB~mag) & [0,6] & [0,7]  \\
        \hline
        IMF & \citet{chabrier_galactic_2003}  & \citet{kroupa_variation_2001}  \\
        \hline
    \end{tabular}
\end{table}

\begin{figure}[t]
    \centering
    \includegraphics[width=0.9\columnwidth]{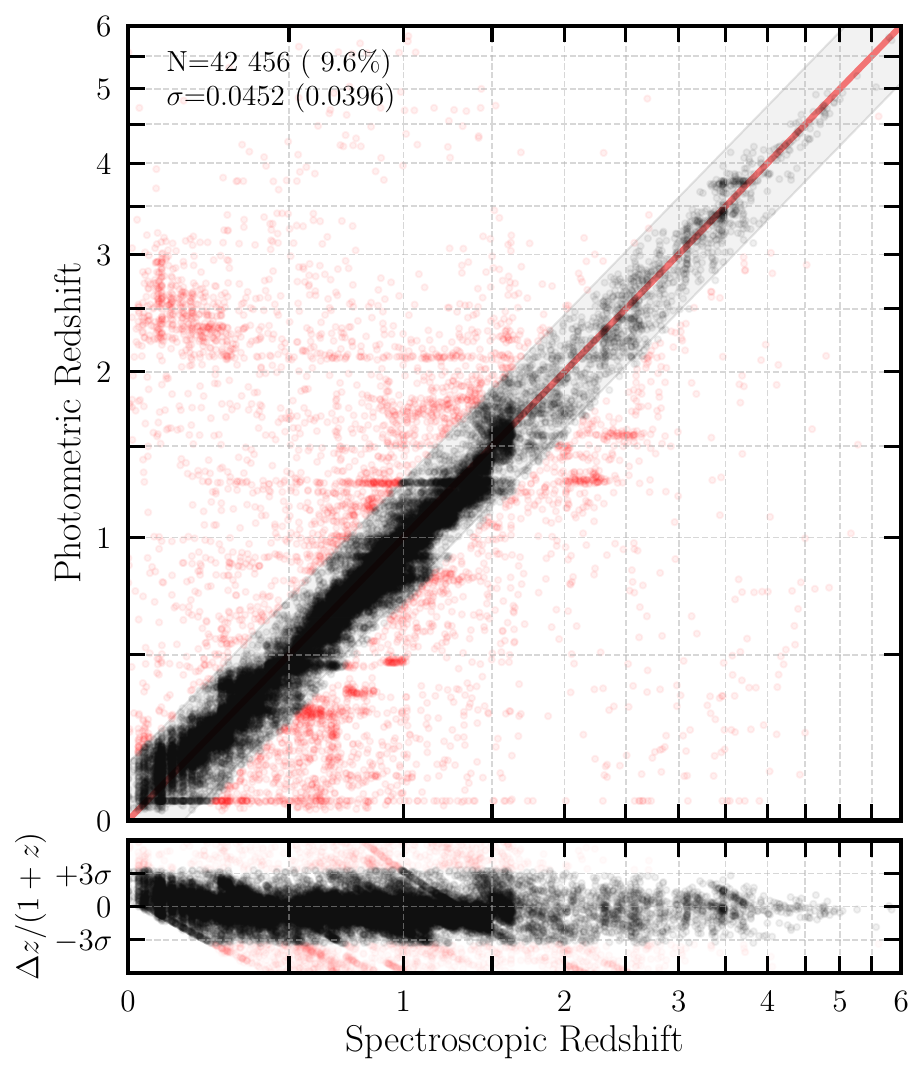}
n\caption{Photometric redshifts that we derive in our work versus the spectroscopic redshifts from the literature available in EDF-N and EDF-F. For a more detailed description of the spectroscopic sample, we refer the reader to \cite{Q1-TP005}. We report an outlier fraction of $9.6\%$ over $\sim 42\,000$ galaxies in $z\in(0,6)$ at all magnitudes. We show the identity as a red continuous line, together with the shaded area delimiting catastrophic outliers, defined as $|z_{\rm phot}-$$z_{\rm spec}|/$$(1+z_{\rm spec})>0.15$. These cases are highlighted in red colour.}
\label{fig:zphot_zspec}
\end{figure}

\section{\label{sc:sedfit} Estimation of photometric redshifts, stellar masses, and other physical properties}

\begin{figure*}[t]
    \centering
    \includegraphics[width=0.95\textwidth]{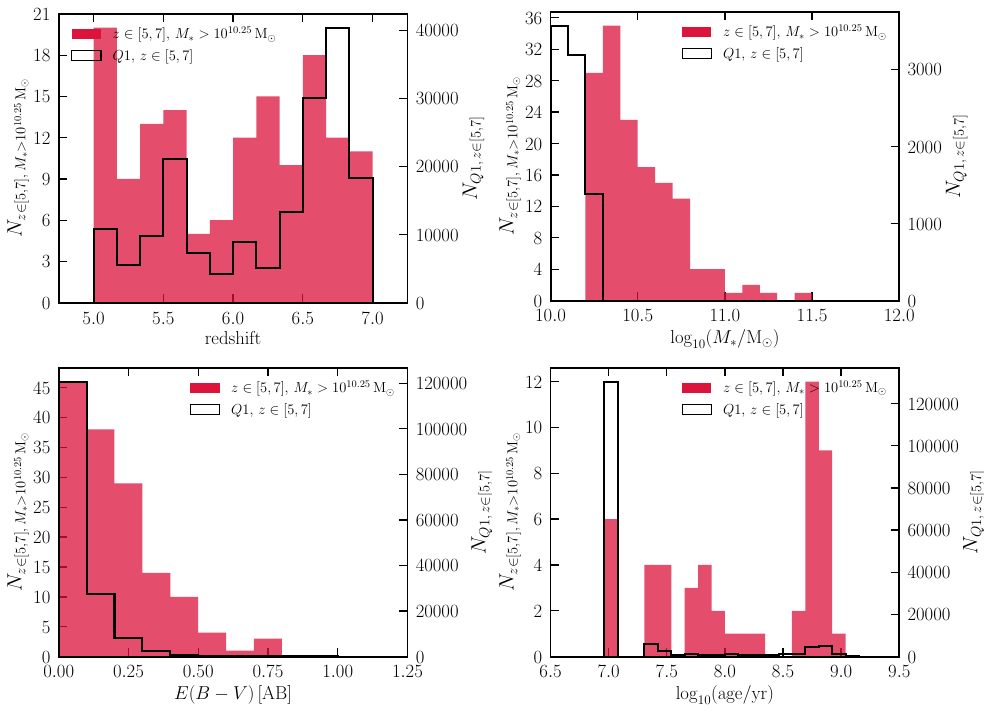} % Replace with your image file
    \caption{Distributions of our massive galaxy sample (red histograms, only for $M_* > 10^{10.25},\mathrm{M_\odot}$) compared to all Euclid Q1 galaxies (black-outlined histograms, not individually visually inspected) in the redshift range $z\in[5,7]$. From left to right and top to bottom, we show the distributions of photometric redshift, stellar mass, best-fit age, and colour excess.}
    \label{fig:photom_prop_pane}
\end{figure*}

\subsection{Photometric redshifts and derived parameters}

We  estimated the photometric redshifts and stellar masses of all NISP-detected galaxies by performing their SED fitting using the code \textsc{LePHARE} \citep{arnouts_lephare_2011}.  We considered a set of stellar population synthesis models from \citet[][hereafter \textsc{BC03}]{bruzual_stellar_2003}, including a single stellar population and a series of exponentially declining star-formation histories with $\tau =$~0.01, 0.1, 0.3, 1.0, 3.0, 5.0, 10.0, and 15~Gyr. All models have been constructed for two possible metallicities, namely Solar ($\rm Z_\odot = 0.02$) and sub-Solar ($Z = 0.2 \, Z_\odot = 0.004$). We also included the empirical QSO templates from \citet{polletta_chandra_2006}.

For all \textsc{BC03} models, we considered a Chabrier \citep{chabrier_galactic_2003} initial mass function (IMF). We convolved the spectral models with a reddening law following the prescription of \cite{calzetti_dust_2000} and \cite{leitherer_global_2002}, with colour excess values of $E(B-V)$~$\in [0, 1.5]$ equally distributed in steps of $0.1$ for all \textsc{BC03} models.

In our photometric catalogue to be used as input for \textsc{\textsc{LePHARE}}, we applied zero-point corrections, as they significantly improve the quality of the photometric redshifts (when comparing them to a spectroscopic sample, see Sect. \ref{sc:sedfit}). We derived these zero-point corrections using the code \textsc{EAZY-Py} \citep{brammer_eazy_2008, brammer_gbrammereazy-py_2021} with the  \citep{van_mierlo_no_2023}. We determined corrections to the zero points of all photometric bands by a direct comparison between the photometric and spectroscopic redshifts using an iterative process (using the SFHZ\_13 template set, which has been validated in literature with similar datasets as ours, see \citealp[]{van_mierlo_euclid_2022}).

With all the above, together with the \textsc{The Farmer} photometry, we have established a finely tuned framework for detecting and characterizing the properties of high-$z$ galaxies identified in the Q1 data.

We obtained photometric redshifts for all the sources and checked a subsample of about 42\,000 sources against available spectroscopic redshifts (Fig. \ref{fig:zphot_zspec}). From this diagnostic, we found that the fraction of catastrophic outliers defined as $|z_{\rm phot}-$$z_{\rm spec}|/$$(1+z_{\rm spec})>0.15$ is $9.6 \%$, and the normalized median absolute deviation of the sample is $0.045$.  Although the available spectroscopic redshifts do not extend to $z>5$, the diagnostic shown in Fig.  \ref{fig:zphot_zspec} allows us to control the overall quality of our photometric redshifts. Particularly, it ensures that no low-$z$ sources are systematically assigned photometric redshifts in our range of interest. We inspected the $\chi^2$ values and found them to be consistent with expectations for good-quality fits.

 \textsc{\textsc{LePHARE}} also provides us with stellar masses and other best-fit model parameters such as colour excess $E(B-V)$ and ages. Taking into account the results of \textsc{\textsc{LePHARE}}, we pre-selected a sample of galaxies with stellar masses $M_* > 10^{10.25} \, \rm M_\odot$  at $z\in [5,7]$  and refined it with a series of selection criteria (see Sect.~\ref{sc:sample_selection}). 

In addition, we made use of the code \textsc{BAGPIPES} \citep{carnall_inferring_2018} to independently check our photometric redshifts. This code is built upon Bayesian inference and nested sampling of the parameter space, which is able to fit photometric data making use of a user-defined parameter set. This flexibility and continuous sampling of the parameter-space makes it virtually impossible to fit all our sample with \textsc{BAGPIPES}, due to time constraints. Therefore, we limit our \textsc{BAGPIPES} run to our pre-selected high-$z$ massive galaxy sample.

Table \ref{tab:sedfit_params} quotes the parameters included in the \textsc{BAGPIPES} fit, together with the range of values used as (uninformative, flat) priors. We adopted a configuration similar to \textsc{LePHARE} for most parameters. 
Notably, \textsc{BAGPIPES} implements a more flexible nebular emission modelling using \textsc{CLOUDY} \citep{ferland_2017_2017}. We allowed for a broad range of ionisation parameter values $\logten U \in (-4, -0.5)$, following \citet{navarro-carrera_interstellar_2024}, by expanding  \textsc{BAGPIPES} default grid of nebular models using \textsc{CLOUDY}.

\section{\label{sc:sample_selection} Massive galaxies at the end of the EoR}

% \begin{figure*}[t]
%     \centering
%     \includegraphics[width=0.75\textwidth]{mosaic.png} % Replace with your image file
%     \caption{RIZ false-color image of 25 galaxies from our sample of DAWN-detected massive candidates. Scale is $5x 5"$. \textcolor{red}{\bf To be replaced with new Euclid stamps... }}
%     \label{fig:mosaic_rgb}
% \end{figure*}

With the goal of selecting a robust sample of massive galaxies at $z\in[5,7]$, we applied a series of criteria as follows. We note that, given the depth of the Euclid Q1 data, the selected galaxies will constitute only a lower limit to the total population of massive galaxies at high $z$, as the most dust-obscured objects would remain undetected in a $\HE<24$~mag survey \citep[see e.g.,][]{caputi_spitzer_2015}.

\subsection{Selection function}

We performed an initial selection of candidates taking into account the redshifts and physical properties derived from \textsc{LePHARE}. For our initial selection of candidates we impose:
\begin{enumerate}
    \item Detection in \textit{Spitzer}/IRAC1 and/or IRAC2. Although we do not impose a minimum signal-to-noise ratio in the IRAC bands (see \citealt{EP-Zalesky}),  we inspected visually all our candidates to ensure their robust detections in both IRAC and NISP-stack.
    \item A probability greater than $70\%$ of being in the redshift bin $z\in[5,7]$, as defined by the probability distribution of photometric redshift produced by \textsc{LePHARE}. In other words, $\int_{5}^{7} p(z)\,{\rm d}z \geq 0.7$. This takes into account the full probability distribution and not only the best-fit value.
    \item Stellar mass that satisfies $M_* \geq10^{10.25}\, \rm M_\odot$ (\textsc{LePHARE}). We start with a rather low stellar mass cut, and then we analyse the impact of making it stricter.
    \item No detections above $2\sigma$ past the Lyman limit (HSC-$u \, g$ bands, at $z>5$).
    \item Independent \textsc{BAGPIPES} and \textsc{LePHARE}++ (DAWN catalogue) photometric redshift solutions compatible with our \textsc{LePHARE} run. We require that $| \, z-z_{\textsc{LePHARE}}|/(1+z_{\textsc{LePHARE}}) < 0.15$ where $z_{\textsc{LePHARE}}$ is our \textsc{LePHARE} photo-$z$ determination.
\end{enumerate}

To address the possible contamination by brown dwarfs, which mimic the colours of true high-$z$ galaxies, we included empirical spectra of L,  M, and T stars from the SpeX Prism Library \citep{burgasser_spex_2014} following \cite{van_mierlo_euclid_2022}. We confirmed that none of the secure sources is best fit with a single-star template using \textsc{LePHARE}. Furthermore, the rich ancillary optical and NIR data help us to increase the purity of our sample, as discussed in \cite{van_mierlo_euclid_2022}.

\subsection{Impact of star formation histories and emission line prescription in stellar-mass determinations}

One important concern for our derived stellar masses is that they could be overestimated because of the presence of plausibly prominent emission lines (H$\beta$+[\ion{O}{iii}], H$\alpha$+[\ion{N}{ii}]) in the IRAC1 and IRAC2 for our redshift range of interest, namely $z\in[5,7]$. To investigate the impact of emission lines in our stellar mass determinations, we did the following.

First, we re-ran \textsc{LePHARE} for our massive galaxy candidates, with the already-obtained photometric redshifts fixed, but this time switching off the two IRAC bands. Stellar masses computed without IRAC photometry are on average larger by $0.15 \ \rm dex$ and show a scatter of $0.37\ \rm dex$ when compared to the original ones.

As a second step, we performed the SED modelling using the code \textsc{BAGPIPES}, leaving all parameters (including redshift) free within the intervals reported in Table \ref{tab:sedfit_params}. \textsc{BAGPIPES} plays an important role in indenendently assessing whether \textsc{LePHARE}'s photometric redshifts and stellar masses are robust, and the latter do non result to be systematically overestimated due to the effect of emission lines.

When using a exponential star formation history (SFH), we report a scatter of $0.25~\mathrm{dex}$ in the stellar mass comparison. The \textsc{LePHARE}-\textsc{BAGPIPES} comparison indicates that the latter yields stellar masses that are, on average, lower by $0.13~\mathrm{dex}$. Results are comparable when using a delayed exponential SFH, with a scatter of $0.31~\mathrm{dex}$ and a systematic offset of $0.19~\mathrm{dex}$.  All but one of the $M_* > 10^{11} \ \mathrm{M_\odot}$ candidates remain above the selection threshold when using \textsc{BAGPIPES} stellar masses. The remaining source has a \textsc{BAGPIPES}-derived mass consistent with the \textsc{LePHARE} estimate within the reported scatter.

For both SFHs, none of our candidates has a \textsc{BAGPIPES} best-fit stellar mass significantly different from the  fiducial values. We did not discard any massive galaxy candidate on the basis of its stellar mass.

We conducted additional \textsc{BAGPIPES} runs employing composite star formation histories. Specifically, we allowed for an older stellar population alongside a recent burst of star formation. However, upon analysing the resulting posterior distributions, we observed that the photometric dataset lacks sufficient constraints for composite SFH models, primarily due to the substantial number of free parameters relative to a single component.
 
After all these tests, our sample of  $M_* >10^{10.25}\, \rm{M_\odot}$ galaxies at $z\in[5,7]$ contains 145 objects, including 5 galaxies with $M_*>10^{11} \, \rm{M_\odot}$. These are \textit{Euclid} galaxies with a counterpart in the DAWN catalogue and which are detected in IRAC. The following analysis only refers to them.

 \begin{figure}[t]
    \centering
    \includegraphics[width=1\columnwidth]{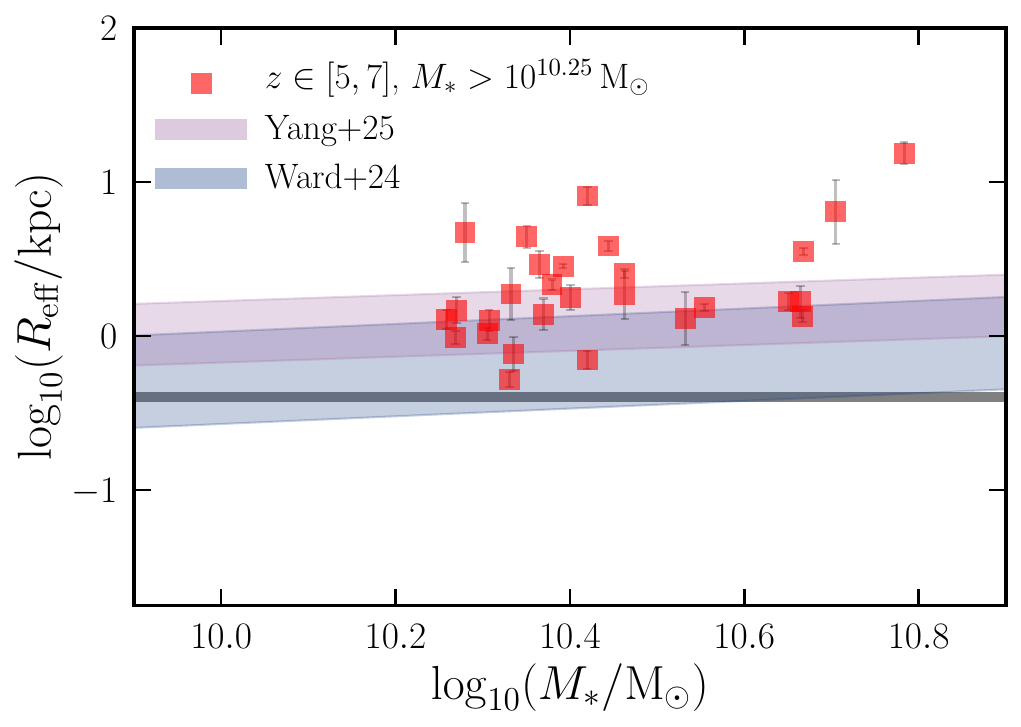}
    \caption{Size–stellar mass relation for VIS-detected galaxies in our sample (36 objects), showing the Sérsic effective radius from \IE modeling as a function of stellar mass. Our high-mass galaxies are represented by red squares. Observational relations from recent JWST-based studies are included as shaded regions: blue for \citet{ward_evolution_2024} and purple for \citet{yang_cosmos-web_2025}. The minimum effective radius (\IE) at the average redshift of our sample is shown as a gray line.}
    \label{fig:size_mass}
\end{figure}

\begin{figure*}[t]
    \centering
    \includegraphics[width=\textwidth]{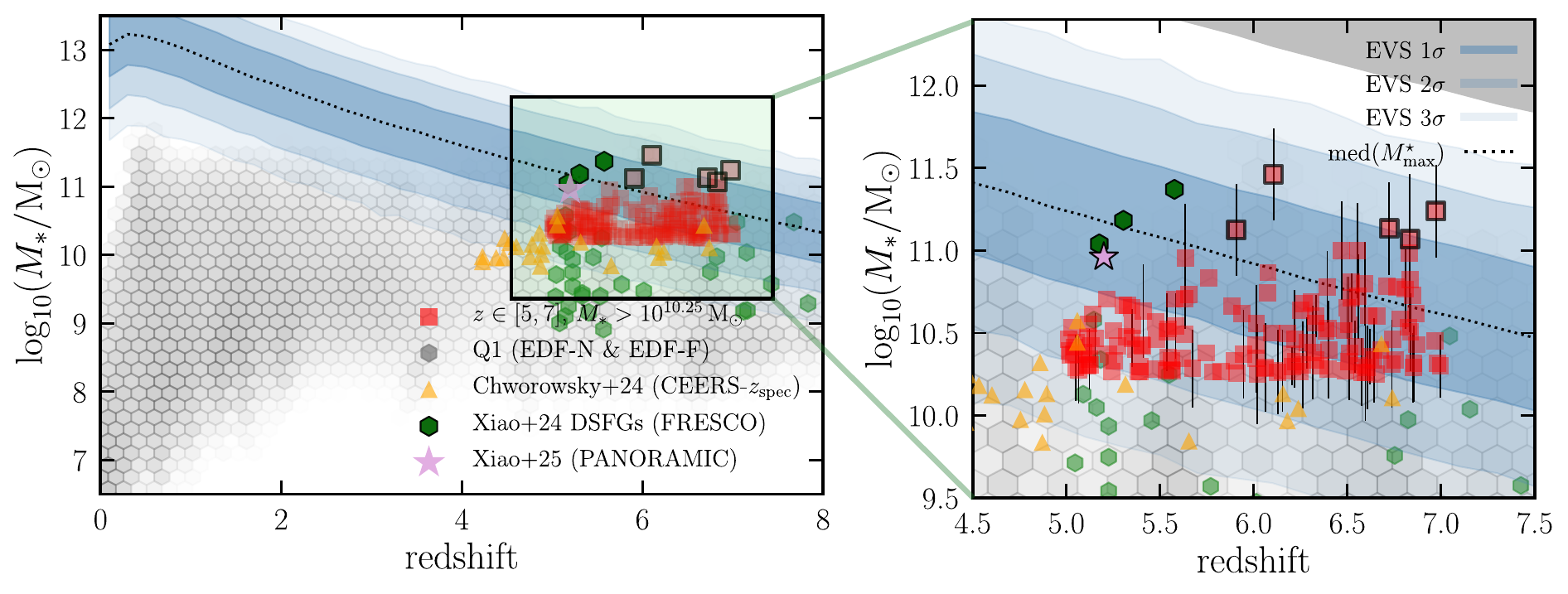} % Replace with your image file
    \caption{\emph{Left panel:} Stellar mass ($M_* $) versus photometric redshift. Our massive galaxy candidates are shown as red squares, overlaid on a gray hexagonal density plot representing the Q1 sample. The 5 candidates above $M_* >10^{11}\, \rm{M_\odot}$ are shown with a black outline. For comparison, we include massive galaxies reported in the literature: orange triangles from \citet{chworowsky_evidence_2024}, green hexagons from \citet{xiao_accelerated_2024} (with black outlines highlighting the three most massive objects S1, S2, and S3) and a pink star from \citet{xiao_panoramic_2025}.
    \emph{Right panel:} Zoom-in of the left panel, focusing on our massive galaxy candidates shown as red squares. Confidence intervals from the Extreme Value Statistics (EVS) model \citep{lovell_extreme_2023}, computed for a survey area comparable to ours, are shown as blue shaded regions, by assuming a lognormal distribution of the star-formation efficiency (SFE). The gray shaded area at the top represents the absolute upper limit under the assumption of 100\% SFE.
    \label{fig:hmf_panel}}
\end{figure*}

\subsection{Physical properties of massive galaxies}

The properties of all galaxies at $z\in[5,7]$, along with our massive galaxy subsample within that redshift range, are shown in Fig.~\ref{fig:photom_prop_pane}. Most massive galaxies in our sample have stellar masses below $10^{11} \, \rm M_\odot$. While galaxies with higher stellar masses do exist, they represent only a few percent of the sample and are predominantly located at $z \in [6,7]$.

About a half of the massive galaxies have best-fit ages close to the age of the Universe at their redshifts, which suggests that they formed at $z\gg7$. This is in contrast to less massive galaxies at those redshifts, most of which prefer very young ages. In addition, the most massive galaxies display a range of $E(B-V)$ values up to 0.75. Instead, the majority of lower mass galaxies at  $z\in[5,7]$ are dust-free. 

When analysing the \textsc{BAGPIPES} results assuming the same SFH but different samplings of age, metallicity, and dust extinction, we find that some of the oldest galaxies instead display intermediate ages and higher $E(B-V)$ values compared to those derived with \textsc{LePHARE}. In particular, most of the old ages ($> 10^{8.5} ~ \rm yr$) recovered by LePhare are distributed more uniformly between $10^{7.5} - 10^{8.5} ~ \rm yr$. This is tied to the distribution of dust extinction (using the same \citealp[]{calzetti_dust_2000} law), where in turn, BAGPIPES recovers slightly higher values of $E(B-V)$. The median $E(B-V)$ is $0.1 ~ \rm mag$ for LePhare and  $0.2 ~ \rm mag$ for BAGPIPES.

This degeneracy is well known in the literature \citep{papovich_stellar_2001, tacchella_stellar_2022}. Nonetheless, the quality of the best-fits measured from their $\chi^2$, is not (statistically) different when comparing both solutions. Furthermore, the stellar mass estimates from both codes are consistent within the uncertainties in most cases. Additional photometry would be required to improve the coverage of the rest-frame optical regime and of stellar population age–sensitive features such as the Balmer break \citep{vikaeus24}.

For the subsample of galaxies detected in VIS (24$\%$ with a S/N~$>5$), we utilize morphological parameters derived from Sérsic modelling in the Q1 Euclid catalogue to study the size–stellar mass relation. We choose to use VIS imaging, as its resolution allows us to probe radii as small as $0.45$~kpc at $z = 6$ (FWHM of 0\farcs16; \citealp{EuclidSkyVIS}), in contrast to NISP, which has a larger minimum resolvable radius of $1.15$~kpc (FWHM of 0\farcs4; \citealp{EuclidSkyNISP}).

In light of Fig.~\ref{fig:size_mass}, the VIS-detected subset of our massive sample shows moderate agreement with recent size–mass calibrations at similar redshifts by \citet{ward_evolution_2024} and \citet{yang_cosmos-web_2025}, although neither study probes the stellar mass regime covered here. We note the presence of a minority of galaxies exhibiting slightly larger sizes than predicted by these relations.
%%%%%%%%%%%%%%%%%%

\begin{figure}[t]
    \centering
    \includegraphics[width=1\columnwidth]{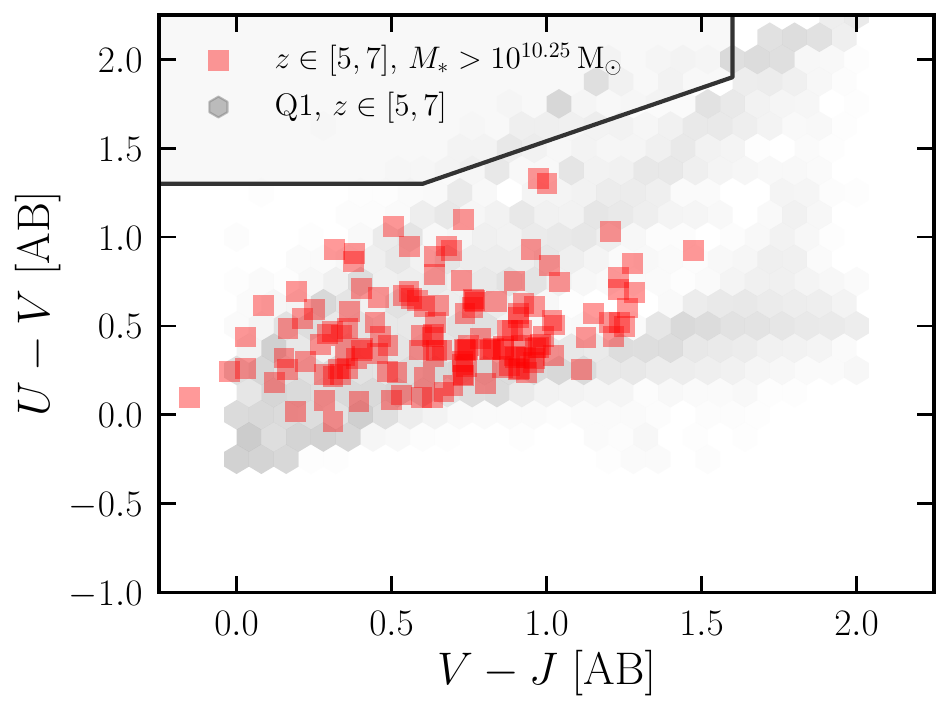}
    \caption{\textit{UVJ} diagram  for the massive galaxy sample (red squares), with the full Q1 population at $z\in[5,7]$ represented as a gray hexagonal kernel density overlay. The galaxy colours for this diagram have been computed from \textsc{LePHARE}-derived rest-frame magnitudes.}
    \label{fig:u}
\end{figure}

\subsection{Results in the context of theoretical predictions}

Recent works have claimed to find an over-abundance of very massive galaxies at $z > 5$ using JWST. This is because the areas probed by JWST surveys are small ($\ll 1 \, \rm deg^2$), so even a few high-$z$ massive galaxies produce a high comoving number density. Some examples of these works are \cite{chworowsky_evidence_2024}, \cite{xiao_accelerated_2024}, \cite{xiao_panoramic_2025}, \cite{akins_two_2023}, and \cite{carnall_jwst_2024}. Their findings are apparently in tension with the predictions of $\Lambda\mathrm{CDM}$-based galaxy formation models  \citep{boylan-kolchin_stress_2023, lovell_extreme_2023}.

Here, we employ the extreme value statistics method by \citet[][hereafter EVS]{lovell_extreme_2023} to assess whether the most massive objects in our sample, given their redshift and the survey area,  are consistent with expectations from standard halo mass functions (HMFs), along with typical baryon and stellar mass fractions used to convert the HMF into a stellar mass function (SMF). For a detailed description of this methodology, we refer the reader to \citet{lovell_extreme_2023}.

Figure~\ref{fig:hmf_panel} (left panel) shows the stellar-mass versus redshift distribution for our massive galaxy sample, placed in the context of the full \textit{Euclid}/DAWN galaxy population. The right panel provides a zoomed-in view of our sample's location, overlaid with the EVS predictions for a survey of comparable area. 

We find no significant tension between our observations and the EVS predictions: only 2 out of 145 galaxies fall outside the $1\sigma$ confidence region. None of our galaxies lie within the forbidden region that would imply star-formation efficiencies of 100\%. In other words, we do not identify any galaxy with an unexpectedly high stellar mass given the survey area of approximately $23~\rm{deg}^2$.

This stands in contrast to previous studies, which report similarly massive galaxies at comparable redshifts, but detected within substantially smaller survey areas. For comparison, Fig.~\ref{fig:hmf_panel} includes a selection of these massive galaxies recently reported using JWST data. Nonetheless, some of these studies \citep{xiao_accelerated_2024} target heavily dust-obscured massive galaxies, whose detectability is limited when using \textit{Euclid}+IRAC data, as in our analysis.

Taken at face value, the surface density of massive galaxies with $z\in[5,7]$ and $M_*  > 10^{10.25} \, \rm{M_\odot}$ in our sample is approximately $6.3~ \rm{deg^{-2}}$. This value decreases to about $0.2~\rm{deg^{-2}}$ for galaxies with $M_*> 10^{11} \, \rm{M_\odot}$. However, not that we are likely missing the most heavily dust-obscured objects due to the limited depth of the NISP bands \citep[see e.g.][]{caputi_spitzer_2015}.

\section{The SFR-$M_*$ plane}
\label{sec:sfrmst}

\subsection{\label{ssc:sfrs_calculation}Star-formation rates from the rest-frame UV}

We derived the star-formation rate from the rest-frame UV luminosity of each galaxy, based on the observed photometry in the filter encompassing the rest-frame UV,  following the methodology detailed in \cite{navarro-carrera_burstiness_2024}. More precisely, we measured the UV luminosity density at $\lambda \in [1500  \, \rm  \AA, 2800  \, \rm \AA]$ by selecting the closest corresponding observed filter based on the best redshift solution for each source. We corrected for dust extinction using the \citet{calzetti_dust_2000} reddening law, taking into account the colour excess values derived from the best-fit SED.

We then estimated the dust-corrected UV-SFR by converting the UV luminosity density at $\lambda \in [1500 \rm{\AA}, 2800 \rm{\AA}]$, following the prescription of \citet{kennicutt_star_1998} and the metallicity-dependent calibrations by \citet{theios_dust_2019}.

Figure \ref{fig:u} shows the (rest-frame) \textit{UVJ} diagram for all \textit{Euclid} galaxies at $z\in[5,7]$ in EDF-N and EDF-F, with our selected massive galaxy candidates highlighted. We see that all these massive galaxies lie in the star-forming region, with none of them present in the passive-galaxy wedge. This result is not surprising: the depth of the Euclid Q1 data does not allow for the identification of passive galaxies. Notably, a small fraction of the lower stellar-mass galaxies at $z\in[5,7]$ do appear in the passive wedge. The analysis of these sources is beyond the scope of this paper, but we note that they most likely are dusty star-forming galaxies whose colours mimic those of passive galaxies.  In Sect.~\ref{ssc:sfrs_seq}, we examine the location of our star-forming massive galaxy sample in the SFR versus stellar-mass plane.

\subsection{\label{ssc:sfrs_seq} Starbursts and main-sequence galaxies  in the SFR-$M_*$ plane}

\begin{figure}[t]
    \centering
    \includegraphics[width=1\columnwidth]{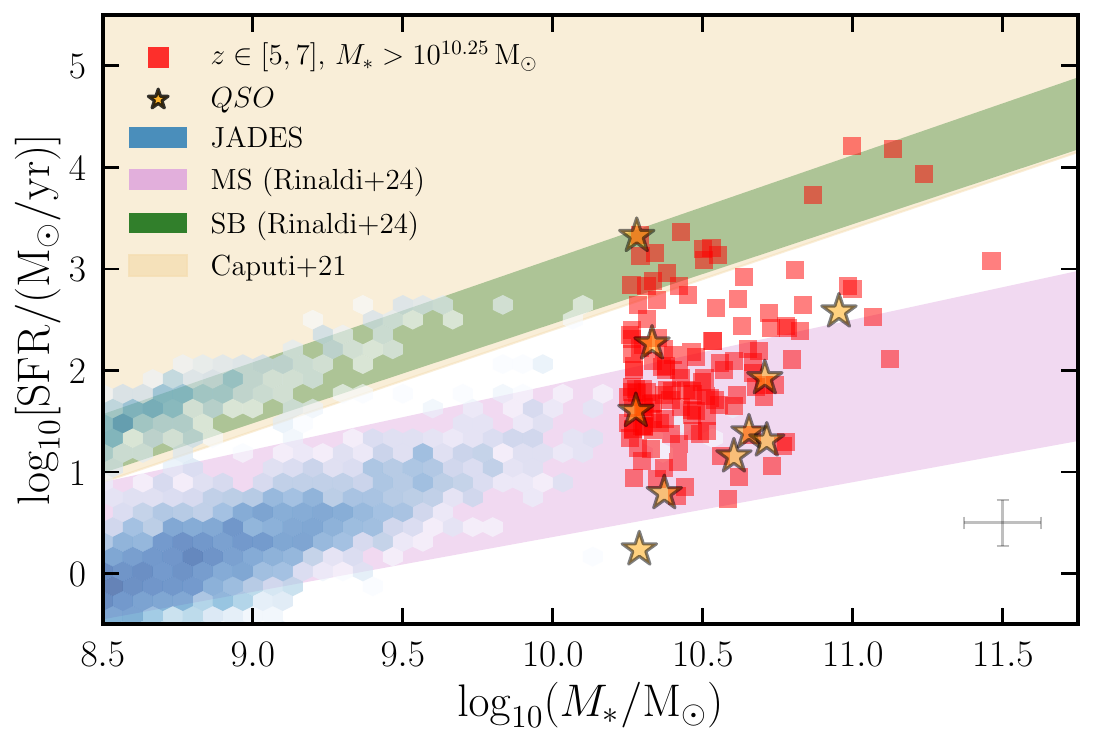}
    \caption{SFR-$M_*$ plane depicting our massive galaxy candidates as red squares.  \textsc{LePHARE}-selected QSOs from our sample are shown as orange, black-outlined stars. The distribution of all  galaxies with $z\in[5,7]$ from the JADES sample \citep{navarro-carrera_burstiness_2024} is shown as a blue hexagon density plot. Green and purple solid areas depict the best fit for SB and MS from \cite{rinaldi_emergence_2025} for $z\in[5,7]$, respectively. Finally, the light coloured orange area represents the  SB semi-plane empirically defined by \citet{caputi_star_2017, caputi_alma_2021}, namely $\rm sSFR_{UV} > 10^{-7.6} \, \rm yr^{-1}$. All SFRs are derived from rest-frame UV  luminosities, as described in Sect. \ref{ssc:sfrs_calculation}.}
    \label{fig:sfr_m}
\end{figure}

Figure \ref{fig:sfr_m} shows the location of our massive galaxies, as well as all other galaxies at $z>5$,  in the SFR-$M_*$ plane. We see that about 86$\%$ (124 out of 145) of our massive galaxy candidates are found around the so-called star-formation main sequence \citep{speagle_highly_2014, bisigello_analysis_2018, rinaldi_emergence_2025}, while the remaining ones appear to have significantly higher SFR values and are thus located in the starburst cloud \citep{caputi_star_2017}. The secondary \textsc{BAGPIPES} run (see Sect. \ref{sc:sedfit}) confirms the presence and relative abundance of starburst galaxies and is consistent with the  \texttt{LePhare} results.

Even if a minority, the presence of massive galaxies in the starburst cloud is very puzzling. Most of these galaxies have best-fit young ages 
$\in (10^7 \, \rm yr, 10^8 \, \rm yr)$ and significant dust extinction ($A_V\in[1.5,3.0]$), as many starbursts do. Actually, most low-stellar mass galaxies are starbursts at high $z$ \citep{rinaldi_galaxy_2022, rinaldi_emergence_2025}, but this possibility is less obvious for massive galaxies:  the empirically defined starburst limit, i.e. $\logten(\rm sSFR/yr^{-1})  > -7.6$ \citep{caputi_alma_2021}, implies a stellar mass doubling time $< 40 \, \rm Myr$. So it is difficult to explain how such massive galaxies could be formed in such fast star-formation episodes. Alternatively, as we discuss in Sect.~\ref{ssc:scrutiny_starbursts}, their presence in the starburst region could suggest that their nature is more complex and the SFRs are overestimated.

\subsection{\label{ssc:scrutiny_starbursts} Scrutiny of the nature of massive starbursts}

Here we investigate whether the SFRs derived for the massive galaxies that appear in the starburst cloud are correct, or whether their rest-frame UV luminosities could have (at least partly) a different origin, leading to an overestimation of the SFRs. Indeed, in a study of H$\alpha$ emitters in the COSMOS field at $z\in (4,5)$, \citet{caputi_spitzer_2015} found that most of their massive galaxies lying in the starburst cloud were in fact hosting AGN, as determined via their X-rays or \textit{Spitzer} $24 \, \rm \mu m$ detections. Thus, they concluded that their derived SFRs were most likely overestimated.

For this purpose, we first checked whether any of our sources has a best-fit SED  (from \textsc{LePHARE}) with a QSO template rather than a normal galaxy template (\textsc{CHI\_QSO}$<$\textsc{CHI\_GAL}). We found 10 cases where this happens. For all these galaxies, the best-fit QSO template  also produces a $z>5$ solution. However, only one of these galaxies lies in the starburst cloud. Additionally, we report that none of the galaxies in our sample falls in the (Q1) \Euclid colour-selected sample of AGN produced by \cite{Q1-SP023}.

To further investigate the nature of our candidates, we performed a cross-match with the red colour-selected galaxies from \citet{Q1-SP016}. We find that 10 of our candidates satisfy the red colour criterion defined in their study, namely NISP \HE--IRAC2~$>$~2.25. Among these, a total of two objects are found in common between the two samples. Nonetheless, our analyses differ substantially in terms of photometric extraction, SED-fitting procedures, and the criteria adopted for sample cleaning and selection.

In addition, we cross-matched our sample with AGN catalogues based on ancillary data at different wavelengths. We do not find any match within the X-ray catalogues of \citet{hasinger_rosat_2021}  and \citet{krumpe_chandra_2015}, nor with the WISE AGN catalogue of \citet{secrest_identification_2015}. We note, however, that both these X-ray and WISE mid-IR catalogues are very shallow, so the non-detection of our massive galaxy candidates cannot exclude the presence of nuclear activity.

Finally, we examine the optical–infrared colour selection criteria proposed by \citet{chehade_two_2018} and \citet{shin_infrared_2020}, which use combinations of optical and near-infrared colours (e.g., W1–W2–$z$ and $r$–$i$–$z$–\JE, respectively) to identify high-redshift quasars in wide-area surveys such as WISE \citep{wright_wide-field_2010} and ELAIS-N1 \citep{sabater_lofar_2021}. None of our sources simultaneously satisfies all the proposed selection criteria. Moreover, these colour-colour diagrams confirm that our galaxies do not occupy regions typically associated with brown dwarfs, further supporting the reliability of our sample.

We also investigate the possibility that outshining by young stellar populations \citep[e.g.,][]{narayanan_outshining_2024, harvey_behind_2025} may lead to the identification of massive starbursts. To test this, we use \textsc{BAGPIPES} to fit composite star-formation histories, consisting of an old stellar component combined with a recent burst. Given the current photometric coverage, we do not find a significant improvement in the fit quality. However, this result may indicate that improved photometric coverage or spectroscopic observations might be required to better constrain the star-formation histories of these galaxies.

\section{\label{sc:conclusions} Summary and conclusions}

We identified 145 galaxies with stellar masses $M_*>10^{10.25} \, \rm M_\odot$, including 5 with  $M_* >10^{11} \, \rm M_\odot$ at $z\in[5,7]$, over the 23~deg$^{2}$ of the Euclid Q1 EDF-N field  with DAWN ancillary data.  This makes for a rather low surface density $\approx 6.3 \, \rm deg^{-2}$ ($\approx 0.2 \, \rm deg^{-2}$ for $M_* >10^{11} \, \rm M_\odot$ galaxies). These values should be considered lower limits, given the depth of the Euclid Q1 images. Indeed, \cite{caputi_spitzer_2015} performed an analysis of deeper near-IR data (albeit in a much smaller area within the COSMOS field) and suggested that the number density of such massive galaxies at $z\in[5,6]$ could at least be twice as large.

In all cases, our massive galaxies can be explained by being formed in dark matter haloes whose star-formation efficiencies are compatible with common values expected from theory. So our galaxies do not violate the predictions of $\Lambda$CDM-based galaxy formation models.

Our massive galaxy candidates have some noteworthy properties: their best-fit SEDs span a wide range of colour excess, namely $E(B-V) \in [0,0.75]$. This means that, in spite of the bright $\HE<24$ cut characterizing the Q1 galaxy sample, some systems have significant dust attenuation, which is uncommon amongst the lower stellar mass sources. This trend is similar to what has been found at lower redshifts \citep[e.g.,][]{reddy_goods-herschel_2012}. In addition, half of the most massive galaxies appear to be as old as the Universe at their redshifts, which suggests that they were formed at very early cosmic times.

The vast majority of our galaxies lie on the star-formation MS on the SFR-$M_*$ plane, but there is a minority of sources which appear in the starburst region or in the transition zone, suggesting that they could be entering or leaving the starburst cloud. The position of these galaxies in the SFR-$M_*$ plane is puzzling, because massive galaxies are not expected to be starbursts (as this would require that they assemble their large stellar masses within a few $10^7$~yr). At low $z$, massive starbursts are very rare \citep{rodighiero_lesser_2011}. At $z\in[4\text,5]$, the known massive galaxies that appear on the starburst cloud are actually AGN candidates \citep{caputi_star_2017}, suggesting that their SFR could be overestimated. For our \textit{Euclid} massive galaxy sample, we have little evidence that nuclear activity could be hosted in them, although this could well be due to observational limitations. The X-ray and mid-IR catalogues available in the EDF-N and EDF-F are shallow, so the non-detection of our sources remains inconclusive. Further follow up of our massive galaxies is necessary to better understand their nature and confirm their high stellar masses and SFR.

%-------------------------------------------------------------------
\begin{acknowledgements} \AckQone

  We thank Adam Carnall for useful discussions. RNC and KIC
acknowledge funding from the Dutch Research Council (NWO) through the award of the Vici Grant
VI.C.212.036.  
  \AckEC
\end{acknowledgements}

%-------------------------------------------------------------------
% \appendix
% \section{\label{sc:apendix} Products used for visual inspection of the massive candidates.}

% \textcolor{red}{To be updated with new version of figures...}

% \begin{figure*}
%     \centering
%     \includegraphics[width=0.8\textwidth]{dawn.jpg} % Replace with your image file
%     \caption{}
%     \label{fig:sed_dawn}
% \end{figure*}

%
% Here comes the reference list, generated via bibtex from
% the bibfile AandA.bib
%

\bibliography{Euclid, Q1, references2}

\appendix
\section{Source Detection Parameters \label{sec_app:detection}}
The DAWN catalogue was generated using the NISP-stacked image as the detection image, with source extraction performed using \texttt{SEP}. The main detection parameters are summarized in Table~\ref{tab:detection}. In brief, a source is defined as a group of at least 10 connected pixels, each exceeding $5\times$ the global background RMS. A Gaussian filter is applied to model and subtract the background across the image. Source deblending is carried out using the parameters listed in Table~\ref{tab:detection}, ensuring robust separation of neighbouring or partially overlapping objects.

\begin{table}[ht]
    \centering
    \caption{\label{tab:detection} SEP parameters for source detection.}
    \begin{tabular}{ll}
        \hline\hline
        \textsc{Name}    & \textsc{Value} \\
        \hline
        THRESH           & 5 \\
        MINAREA          & 10 \\
        CLEAN            & False \\
        FILTER\_KERNEL   & \texttt{gauss\_1.5\_3x3.conv} \\
        FILTER\_TYPE     & matched \\
        DEBLEND\_NTHRESH & 32 \\
        DEBLEND\_CONT    & $10^{-5}$ \\
        \hline
    \end{tabular}
\end{table}

\label{LastPage}
\end{document}